\documentclass[%
reprint,
superscriptaddress,
showpacs,
amsmath,amssymb,
aps,
]{revtex4-2}
\usepackage{graphicx}% Include figure files
\usepackage{dcolumn}% Align table columns on decimal point
\usepackage{bm}% bold math
\usepackage{CJKutf8}
\usepackage{color}
\usepackage{amssymb}
\usepackage{amsfonts}
\usepackage{palatino}
\usepackage{esint}
\usepackage[colorlinks,urlcolor=blue,linkcolor=blue,citecolor=blue,anchorcolor=blue]{hyperref}
\makeatletter

\newcommand{\Rmnum}[1]{\expandafter\@slowromancap\romannumeral #1@}

\makeatother
\begin{document}

\begin{CJK*}{UTF8}{gbsn}

\preprint{APS/123-QED}

\title{Thresholdless corner vortex solitons in fractal Sierpi\'nski topological insulators}

\author{Yiqi Zhang}
\email{zhangyiqi@xjtu.edu.cn}
\affiliation{Key Laboratory for Physical Electronics and Devices, Ministry of Education, School of Electronic Science and Engineering, Xi'an Jiaotong University, Xi'an 710049, China}

\author{Alexander V. Kireev}
\affiliation{Institute of Spectroscopy, Russian Academy of Sciences, Troitsk, Moscow 108840, Russia}
\affiliation{Moscow Institute of Physics and Technology (National Research University), Dolgoprudny, 141701, Russia}

\author{Victor O. Kompanets}
\affiliation{Institute of Spectroscopy, Russian Academy of Sciences, Troitsk, Moscow 108840, Russia}

\author{Sergey Y. Alyatkin}
\affiliation{Skolkovo Institute of Science and Technology, Moscow, Territory of innovation center ``Skolkovo'', Bolshoy Boulevard 30, bld. 1, 121205, Russia}

\author{Nikita S. Kostyuchenko}
\affiliation{Institute of Spectroscopy, Russian Academy of Sciences, Troitsk, Moscow 108840, Russia}
\affiliation{Quantum Technology Centre, Faculty of Physics, M. V. Lomonosov Moscow State University, Moscow 119991, Russia}

\author{Sergei A. Zhuravitskii}
\affiliation{Institute of Spectroscopy, Russian Academy of Sciences, Troitsk, Moscow 108840, Russia}
\affiliation{Quantum Technology Centre, Faculty of Physics, M. V. Lomonosov Moscow State University, Moscow 119991, Russia}

\author{Nikolay N. Skryabin}
\affiliation{Quantum Technology Centre, Faculty of Physics, M. V. Lomonosov Moscow State University, Moscow 119991, Russia}

\author{Khalil Sabour}
\affiliation{Moscow Institute of Physics and Technology (National Research University), Dolgoprudny, 141701, Russia}

\author{Alexander A. Kalinkin}
\affiliation{Quantum Technology Centre, Faculty of Physics, M. V. Lomonosov Moscow State University, Moscow 119991, Russia}

\author{Yongdong Li} 
\affiliation{Key Laboratory for Physical Electronics and Devices, Ministry of Education, School of Electronic Science and Engineering, Xi'an Jiaotong University, Xi'an 710049, China}

\author{Sergei P. Kulik}
\affiliation{Institute of Spectroscopy, Russian Academy of Sciences, Troitsk, Moscow 108840, Russia}
\affiliation{Quantum Technology Centre, Faculty of Physics, M. V. Lomonosov Moscow State University, Moscow 119991, Russia}

\author{Pavlos G. Lagoudakis}
\affiliation{Skolkovo Institute of Science and Technology, Moscow, Territory of innovation center ``Skolkovo'', Bolshoy Boulevard 30, bld. 1, 121205, Russia}

\author{Sergey V. Chekalin}
\affiliation{Institute of Spectroscopy, Russian Academy of Sciences, Troitsk, Moscow 108840, Russia}

\author{Yaroslav V. Kartashov}
\email{kartashov@isan.troitsk.ru}
\affiliation{Institute of Spectroscopy, Russian Academy of Sciences, Troitsk, Moscow 108840, Russia}

\author{Victor N. Zadkov}
\affiliation{Institute of Spectroscopy, Russian Academy of Sciences, Troitsk, Moscow 108840, Russia}
\affiliation{Faculty of Physics, Higher School of Economics, Moscow 105066, Russia}

\date{\today}% It is always \today, today,
%  but any date may be explicitly specified

\begin{abstract}
\noindent
Quantized vortices are ubiquitous in physics, spanning superconductivity, astrophysics, superfluid condensed matter systems, and nonlinear optics. Yet embedding vorticity into topologically protected nonlinear states has remained a major challenge, with all previously observed corner solitons in higher-order topological insulators (HOTIs) exhibiting only trivial phase distributions. Here, we report on the first realization of stable topological corner vortex solitons in a photonic fractal HOTI. Using an array of laser-written waveguides in the shape of Sierpi\'nski gasket with a controllable distortion, we design linear topological vortex modes, from which nonlinear corner vortex solitons bifurcate. Moreover, we demonstrate that these solitons exhibit exceptional robustness across a broad power range and, unlike vortex solitons in topologically trivial lattices, form without a power threshold. Our results introduce the angular momentum degree of freedom into the physics of topological corner modes, opening prospects for topologically protected vortex-based photonics.
\end{abstract}

\maketitle

\end{CJK*}

\section{Introduction}

Topological materials represent a fundamentally new phase of matter, characterized by a qualitatively distinct behavior of bulk and boundary excitations~\cite{lu.np.8.821.2014, ozawa.rmp.91.015006.2019, zhang.nature.618.687.2023, leykam.nrp.8.55.2026}. Although insulating in the bulk, these materials support the propagation of exceptionally robust in-gap excitations at their edges that are protected by the material's band topology. Such topological excitations can carry information and energy, enabling robust transmission along the edges of material even in the presence of defects, edge deformations or bends. These features align well with modern technologies of information encoding in the phase of propagating fields, including the use of topological charges of nested vortices. However, embedding the vorticity into topological states remains elusive. The properties and localization of topological states are intrinsically tied to the symmetry of the bulk~\cite{wieder.nrm.7.196.2022} and edge geometry of the material. Since vortex-carrying states typically occupy multiple sites of the system, they are difficult to harmonize with specific symmetry of topological edge states. Moreover, the nature of topological insulators imposes fundamental constraints on the dimensionality and spatial extent of the topological states that can emerge. For example, in conventional topological insulators, the effective dimensionality of the edge states is only by one lower than that of the bulk and they are extended in the direction along the edge~\cite{lu.np.8.821.2014, ozawa.rmp.91.015006.2019, zhang.nature.618.687.2023}. Higher-order topological insulators (HOTIs) extend this paradigm by supporting corner-localized modes whose effective dimensionality is reduced by at least two relative to the bulk~\cite{xie.nrp.3.520.2021, lin.nrp.5.483.2023}. Thus, two-dimensional HOTIs~\cite{banalcazar.science.357.61.2017, serra.nature.555.342.2018, noh.np.12.408.2018, xie.prl.122.233903.2019, chen.prl.122.233902.2019, hassan.np.13.697.2019, li.np.14.89.2020, benalcazar.prl.128.127601.2022} can host effectively zero-dimensional corner states.

As a universal physical phenomenon, HOTIs have been demonstrated across diverse platforms, including acoustics, photonics, matter-wave and optoelectronic systems \cite{xie.nrp.3.520.2021, lin.nrp.5.483.2023}. Notably, HOTIs have been constructed not only in bulk-periodic structures \cite{peterson.nature.555.346.2018, noh.np.12.408.2018, hassan.np.13.697.2019, li.np.14.89.2020, mittal.np.13.692.2019, chen.prl.122.233902.2019, xie.prl.122.233903.2019, serra.nature.555.342.2018, zhang.np.15.582.2019, peterson.science.368.1114.2020}, but also in aperiodic systems exhibiting discrete rotational symmetry, with the most prominent examples being disclination structures \cite{peterson.nature.589.376.2021, liu.nature.589.381.2021, li.prb.101.115115.2020, wang.nc.12.3654.2021, ren.apl.8.016101.2023, ren.light.12.194.2023, hwang.np.18.286.2024, huang.nano.13.3495.2024} and fractals~\cite{manna.prb.105.L201301.2022, li.sb.67.2040.2022, zheng.sb.67.2069.2022, ren.nano.12.3829.2023, li.light.12.262.2023, eek.prl.134.246601.2025}. However, the creation of topological states with nontrivial phases like vortices is particularly challenging in HOTIs. Specifically, their excitations are strongly localized in the corners, their spectrum usually permits only single-mode states, and their properties are strongly dependent on geometry. As a consequence, only the simplest, single-mode corner states have been observed experimentally in HOTIs, while incorporation of vorticity into such modes remains a fundamental challenge. Meanwhile, the interplay of topology and nonlinearity has attracted intense interest, because nonlinear self-action enables families of self-sustained states that inherit topological protection from their linear counterparts~\cite{kartashov.prl.128.093901.2022,chen.sb.70.1605.2025,du.light.14.296.2025}. In photonic systems, for example, nonlinearity serves as a versatile knob for control of localization, internal structure and evolution of topological states \cite{smirnova.apr.7.021306.2020, szameit.np.20.905.2024, maczewsky.science.370.701.2020, mukherjee.science.368.856.2020, sone.np.20.1164.2024}. Despite theoretical proposals and experimental progress on nonlinear corner states in HOTIs~\cite{kirsch.np.17.995.2021, hu.light.10.164.2021, ezawa.prb.104.235420.2021, zhong.light.13.264.2024, kompanets.am.37.2500556.2025} and in polaritonics \cite{banerjee.prl.124.063901.2020, zhang.ol.45.4710.2020, wu.sa.9.4322.2023, schneider.nano.13.509.2024, bennenhei.acs.11.3046.2024}, all corner solitons observed to date also have exhibited only trivial phase distributions, while self-sustained corner states carrying global vorticity have not been reported. This is connected with the fact that vortex solitons represent excited higher-order states of the system that are typically prone to azimuthal instabilities and that usually require considerable power thresholds for their formation, at least in topologically trivial systems.

Here we demonstrate experimental observation of \emph{stable topological corner vortex solitons} in a photonic fractal HOTI realized as a Sierpi\'nski-gasket array of fs-laser-written single-mode waveguides. Fractal HOTIs host multiple in-gap corner modes with tunable degeneracies, which enables construction of a linear corner vortex modes using a superposition of states from degenerate corner doublet. Crucially, the emergence of such corner doublets stems from fractal structure of the waveguide arrays and is independent of discrete rotational symmetry of the entire Sierpi\'nski-gasket array. To the best of our knowledge, such corner doublets are not found in conventional HOTIs that were widely analyzed previously and that are based on either square (Su-Schrieffer-Heeger-like) or honeycomb lattices~\cite{xie.nrp.3.520.2021}. The spectra of these systems either lack degenerate corner states entirely, or when they do exist, linear combinations of such states do not lead to vortices near the corners of the structure. We show that self-focusing nonlinearity generates continuous families of corner vortex solitons that bifurcate from an in-gap vortex eigenstate, explaining their \emph{thresholdless} formation in the topological regime. Since the localization of such states strongly depends on their position within topological gap, nonlinearity enables precise tuning of the shapes of corner vortex solitons, thereby introducing a desired control into this topological platform. We confirm vorticity directly via phase-resolved interferometry, revealing a persistent phase singularity between the three corner waveguides for $m=\pm1$.

\section{Theoretical modeling}

Propagation of light in waveguide arrays considered here is governed by the dimensionless nonlinear Schr\"odinger equation for light field amplitude $\psi$
\begin{align}
	\label{NLSE}
	i \frac{\partial \psi}{\partial z} = -\frac{1}{2} \nabla^2 \psi -\mathcal{R}(x,y) \psi - |\psi|^{2} \psi,
\end{align}
where $(x,y)$ and $z$ are the transverse coordinates and propagation distance, respectively; ${\nabla^2\equiv\partial_x^2+\partial_y^2}$; the function 
\[
\mathcal{R} (x,y) =  p \sum_{m,n}  e^{- (x-x_{m,n})^2/w_x^2 + (y-y_{m,n})^2/w_y^2}
\]
describes waveguide array composed from waveguides of depth $p$, widths $w_x,w_y$, with centers in $(x_{m,n},y_{m,n})$ points. Stationary solutions of Eq.~(\ref{NLSE}) have the form $\psi= u(x, y)e^{ibz}$ and can be obtained from the equation 
\[
b u = \frac{1}{2} \nabla^2 u + \mathcal{R}(x,y) u + |u|^{2} u,
\]
where $u$ describes the profile of the state and $b$ is its propagation constant (eigenvalue).

To provide the mapping between dimensionless power $U$ of light beams in numerical modeling that is defined as
\begin{equation}
	U=\iint |\psi|^2 dx dy
\end{equation}
and real peak power $P$, we use standard normalizations adopted in our previous works~\cite{zhong.light.13.264.2024,kompanets.am.37.2500556.2025}, where the real field amplitude $\mathcal{E}$ and the dimensionless one are connected as:
\begin{equation}
	\mathcal{E} = \left( \frac{n}{k^2 r_0^2 n_2 } \right)^{1/2} \psi,
\end{equation}
where $k=2\pi n/\lambda $ and $\lambda=800\,\rm nm$ is the central working wavelength, while $r_0=10\, \mu\rm m$ is the characteristic spatial scale to which transverse coordinates $x,y$ are normalized. The unperturbed refractive index of fused silica is ${n=1.45}$. In this case, the relation between $P$ and $U$ is given by
\begin{equation}
	P=\iint r_0^2 |\mathcal{E}|^2  dxdy = \frac{n}{k^2 n_2 } \iint |\psi|^2 dx dy = \frac{n}{k^2 n_2 } U.
\end{equation}
where the nonlinear coefficient $n_2\approx 2.7\times 10^{-20}\, \rm m^2/W$ at the wavelength ${\lambda=800\, \rm nm}$ for fused silica.

\begin{figure*}[t]
	\centering
	\includegraphics[width=\textwidth]{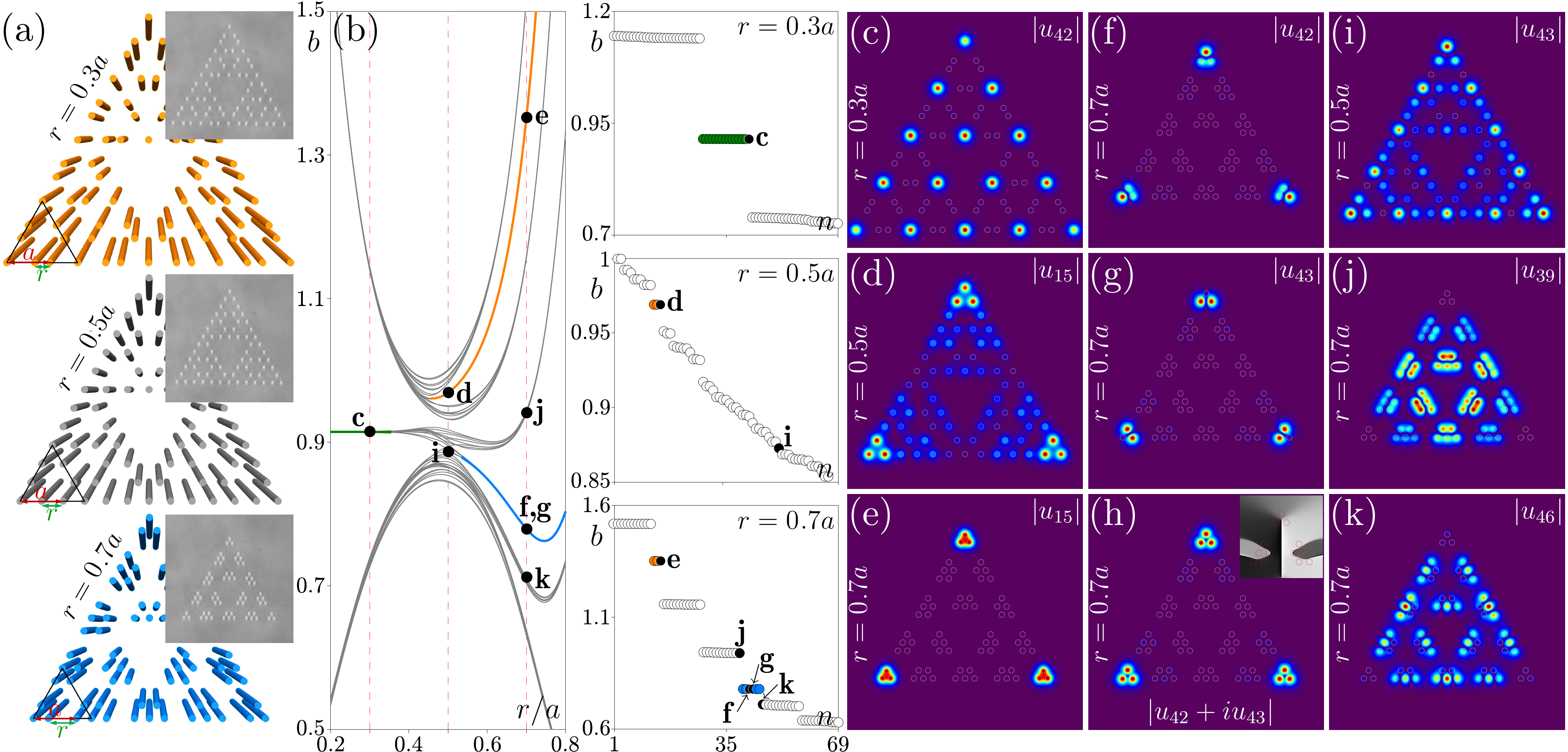}
	\caption{\textbf{Lattices, spectra, and modes.}
		(a) Sketches of the fractal $G_3$ waveguide arrays with different distortion parameters $r$. Triangles indicate $G_1$ elements. The insets show microphotographs of corresponding fs-laser written arrays. (b) Linear spectrum of $G_3$ fractal array as a function of $r$ and eigenvalues $b$ vs mode index $n$ corresponding to three vertical dashed lines in (b). Colored dots correspond to topological states, gray dots to delocalized states. Black dots with letters correspond to representative eigenmodes $u_n$, whose $|u_n|$ distributions (with superimposed on them array contours) are depicted in panels (c-k). Among them are two degenerate corner states $u_{42}$ (f) and $u_{43}$ (g) that can be used to construct corner vortex state ${u_{42}+iu_{43}}$ whose modulus and phase are shown in (h). Panels in (c-k) are shown in ${-25\le x,y \le 25}$. Array parameters are $a=3.4$, $w_x=0.8$, $w_y=0.85$, and $p=3.3$.}
	\label{fig1}
\end{figure*}

\section{Results}
\subsection{Lattice, spectrum, and modes}
We consider third-generation $(G_3)$ Sierpi\'nski gasket fractal arrays created using fs-laser writing~\cite{szameit.jpb.43.163001.2010, skryabin.prap.22.064079.2024,yan.npjn.1.40.2024} as a structure for realization of HOTI. Schematic illustrations and microphotographs of the arrays are shown in Fig.~\ref{fig1}a (see Section S1 in the Supplementary Materials). Self-similar fractal structures characterized by a non-integer effective dimensionality have been established as a perspective platform for investigation of topological phenomena~\cite{manna.prb.105.L201301.2022, li.sb.67.2040.2022, zheng.sb.67.2069.2022, ren.nano.12.3829.2023, zhong.light.13.264.2024, li.light.12.262.2023, eek.prl.134.246601.2025, yang.light.9.128.2020, biesenthal.science.376.1114.2022}. These structures are aperiodic and sometimes considered as lacking a well-defined insulating bulk, while their topological characterization requires real-space invariants~\cite{yang.light.9.128.2020, biesenthal.science.376.1114.2022, li.sb.67.2040.2022, zheng.sb.67.2069.2022}. We introduce controllable Kekul\'e-like distortion (that is also known as Wu-Hu approach~\cite{wu.prl.114.223901.2015}) already into first-generation $(G_1)$ arrays (see Fig.~\ref{fig1}a), characterized by a shift $r$ indicated in $G_1$, while keeping next-nearest spacing $a$ fixed. 
The parameters of arrays entering into function $\mathcal{R}(x,y)$ describing refractive index distribution in such structures are indicated in the caption of Fig.~\ref{fig1}.

Neglecting nonlinearity and using plane-wave expansion method, we obtain the linear spectrum of modes of the fractal array as a function of parameter $r$ displayed in Fig.~\ref{fig1}b (see also the dependencies of all eigenvalues $b$ on mode index $n$ for representative values of ${r=0.3a, 0.5a}$ and $0.7a$ corresponding to dashed lines in $b(r)$ plot). The fractal HOTI hosts multiple in-gap topological modes (colored dots) associated with internal and external boundaries~\cite{eek.prl.134.246601.2025}.

At ${r<0.5a}$, a manifold of nearly degenerate states (green line in Fig.~\ref{fig1}b) appears in the internal and external corners of the structure (see Fig.~\ref{fig1}c). At ${r>0.5a}$ several coexisting topological states in outer corners appear. The orange line in Fig.~\ref{fig1}b corresponds to three-fold degenerate states (mode index ${n=13\sim15}$), where three corner spots are in-phase (Figs.~\ref{fig1}d and \ref{fig1}e). Their localization increases when one moves away from ${r=0.5a}$ point corresponding to equal waveguide spacing. Of particular interest is the blue branch at ${r>0.5a}$ which corresponds to six-fold degenerate modes (index ${n=40\sim45}$) and contains pairs with different structure of spots in array corners (see exemplary states $u_{42}$ and $u_{43}$ in Figs.~\ref{fig1}f and \ref{fig1}g). Their superposition ${u_{42} + i u_{43}}$ forms a linear \textit{corner vortex states} (Fig.~\ref{fig1}h). The phase singularity in such states is positioned between three corner waveguides (inset in Fig.~\ref{fig1}h), and the mode's localization increases with the shift parameter $r$. We stress that the appearance of pairs of degenerate modes residing in the same corner is unique for selected fractal and is not tied to discrete rotational symmetry $\mathcal{C}_3$ of the entire structure (the symmetry axis passes through the center of the array, but not near its corners). At the same time, if one considers three corner waveguides as an isolated subsystem from the rest of the lattice, it would also have discrete rotational symmetry $\mathcal{C}_3$. In accordance with group theory arguments \cite{ferrando.prl.95.043901.2005}, such systems can support vortex modes with topological charges not exceeding ${m=1}$. The examples of delocalized states (associated with gray lines in Fig.~\ref{fig1}b) coexisting with topological ones are shown in Figs.~\ref{fig1}i-\ref{fig1}k. Notice that topological nature of modes from colored branches in Fig.~\ref{fig1}b is confirmed by nonzero values of real-space polarization index for them, as discussed in~\cite{li.sb.67.2040.2022, zheng.sb.67.2069.2022, ren.nano.12.3829.2023, zhong.light.13.264.2024}.

\subsection{Families of corner vortex solitons}

\begin{figure}[h!]
	\centering
	\includegraphics[width=\columnwidth]{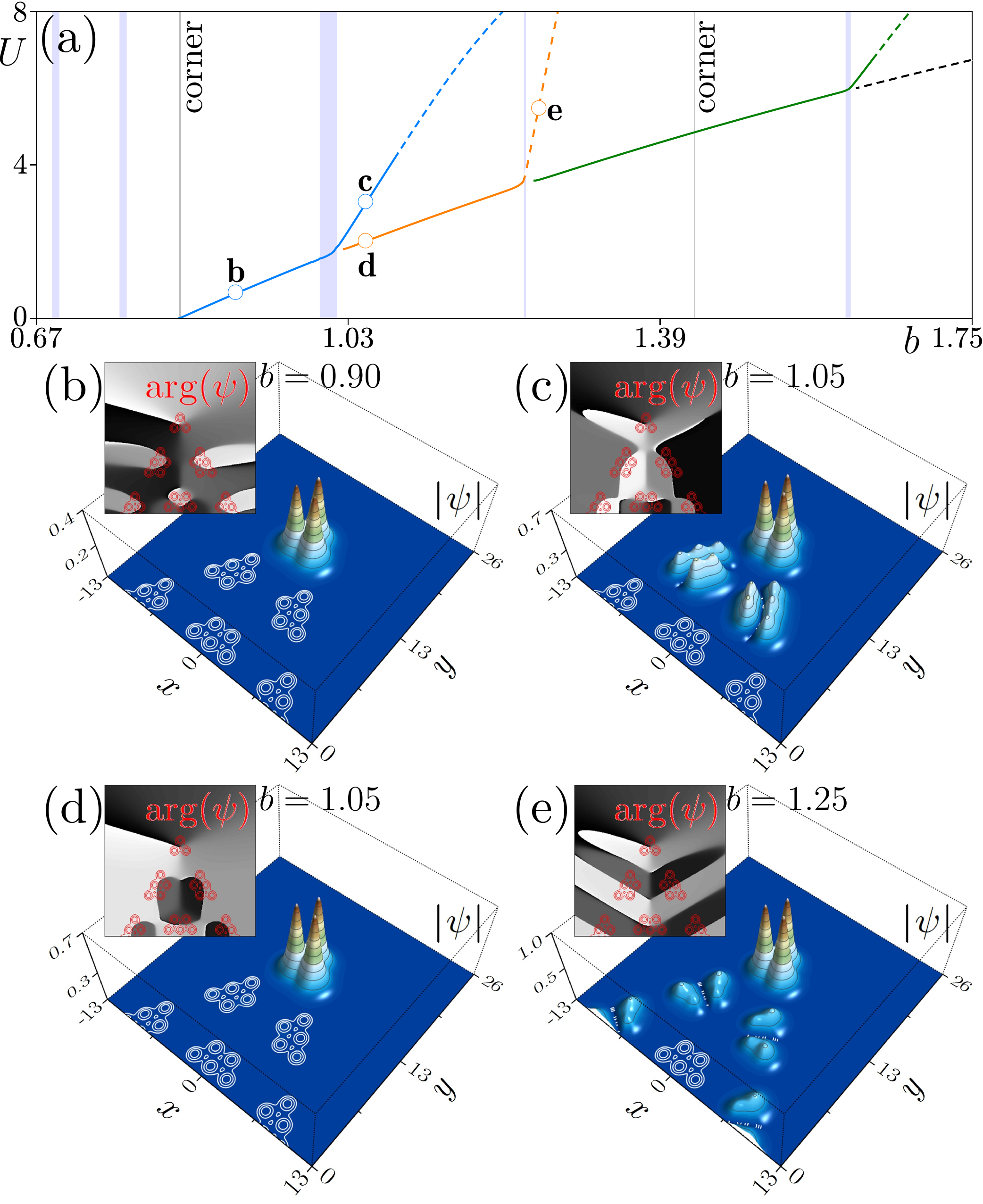}
	\caption{\textbf{Corner vortex soliton family.}
		(a) Power $U$ of corner vortex solitons in fractal array as a function of $b$ at $r=0.7a$. Solid (dashed) lines correspond to stable (unstable) solitons. Purple regions indicate bands of extended states, while gray lines show eigenvalues of linear corner states. (b)-(e) Field modulus and phase (insets) distributions in solitons corresponding to the dots with letters in (a).}
	\label{fig2}
\end{figure}

In nonlinear medium topological corner vortex states give rise to families of corner vortex solitons that can be obtained by the Newton method. They are characterized by power ${U=\iint |\psi|^2 dxdy}$ shown in Fig.~\ref{fig2}a as a function of propagation constant $b$. We emphasize that corner vortex solitons bifurcate from linear vortex states (see blue curve, where ${U\to 0}$ when $b$ approaches eigenvalue of linear vortex at a given $r$) and exist across several spectral gaps. This is the key mechanism underlying \emph{thresholdless} formation: the nonlinear state emerges continuously from an already localized in-gap vortex eigenstate rather than requiring finite-power self-trapping. When $b$ enters into the band of delocalized states, coupling with them occurs resulting in delocalization of corner state and rapid increase of power $U$ (compare soliton profiles in Figs.~\ref{fig2}b and \ref{fig2}c from the same family). Nevertheless, the continuation of the family of vortex solitons can be obtained also in higher-lying gaps (orange and green curves) that in turn couples with delocalized states when power becomes too high (compare profiles in Figs.~\ref{fig2}d and \ref{fig2}e). Thus, one can consider that while the bands of delocalized states locally ``interrupt'' the continuity of the corner vortex soliton families, all such corner vortex soliton families marked by different colors in distinct band gaps do originate from corresponding linear corner states. One can also conclude that nonlinearity introduces tunability in our system and allows for considerable reshaping of self-sustained states even in fabricated structures with a fixed $r/a$ parameter.

The variation of shapes of corner vortex solitons with increase of power strongly depends on chosen value of distortion parameter $r$. Thus, when $r$ is closer to $0.5a$, linear corner vortex states appear much less localized and extend into array. In this case, when one increases propagation constant $b$ (and hence peak power of solution $U$), corner solitons originating from such moderately localized linear corner modes do experience visible reshaping already in the gap -- their localization initially increases with increase of $b$. When propagation constant of such solitons crosses the band of delocalized states, they also exhibit delocalization. In this case one observes nonmonotonic variation of soliton width with increase of $b$, as shown in Section S4 in the Supplementary Materials.

We stress that corner vortex solitons in fractals are stable in broad range of powers, as confirmed by their propagation with added noise (up to $5\%$ in amplitude) in the frames of Eq.~(\ref{NLSE}) over huge distance ${z\sim4000}$, far exceeding sample length. Stable branches in Fig.~\ref{fig2}a are shown with solid lines, unstable -- with dashed lines. Notice stability in finite gaps, and instability in semi-infinite gap (see Section S3 in the Supplementary Materials, where we present also the details of linear stability analysis).

\subsection{Observation of corner vortex solitons}

To realize corner vortex solitons, we used fractal waveguide arrays inscribed in fused silica (unperturbed refractive index ${n=1.45}$ and nonlinear coefficient $n_2\approx 2.7\times 10^{-20}\, \rm m^2/W$ at the wavelength ${\lambda=800\, \rm nm}$) with next-nearest waveguide spacing of ${34\, \mu \rm m}$ (corresponding to ${a=3.4}$) and various shift parameters $r$ from $0.4a$ to $0.8a$. The ${99.1\, \rm mm}$ physical sample length corresponds to $z\approx87$ in our theoretical model (characteristic diffraction length of ${1.14\, \rm mm}$ is used in normalization of $z$ for beam width of ${10~\mu\rm m}$). The evaluated refractive index modulation depth in array ${\delta n \sim 3.7\times 10^{-4}}$ corresponds to ${p=3.3}$ adopted in simulations. The implemented ``multi-scan'' writing technology (see Section S2 in the Supplementary Materials) allows to create waveguides supporting nearly circular modes (even though the cross-sections of such waveguides in microphotographs in Fig.~\ref{fig1}a may look slightly rectangular). This results in a nearly isotropic coupling between the waveguides. In simulations, the couplings in such a system are accurately captured by considering just minimal ellipticity of waveguides: ${w_x=0.8}$ and ${w_y=0.85}$. Namely such minimal ellipticity guarantees that modes giving rise to corner vortex states remain practically degenerate in a broad interval of distortion parameters $r$, see Fig.~\ref{fig1}b. To efficiently excite corner vortex solitons we used a spatially structured pulsed laser excitation ($580\,\rm fs$ pulses of variable energy $E$ obtained from  $1\,\rm kHz$ Ti:sapphire laser system, which provides high peak powers $P$ up to several $\rm MW$). Using the reflective spatial light modulator (SLM), we created three-spot pattern with properly designed phases $2\pi m(\kappa-1)/3$, where $m=\pm 1$ is the topological charge, and ${\kappa=1,2,3}$ is the index of corresponding laser beams. The beams of equal intensities with desired phases are then focused onto three corner waveguides. We note that implemented structured excitation ensures a strong overlap with desired corner vortex state (see Section S2 in the Supplementary Materials for details).

\begin{figure}[htbp]
	\centering
	\includegraphics[width=\columnwidth]{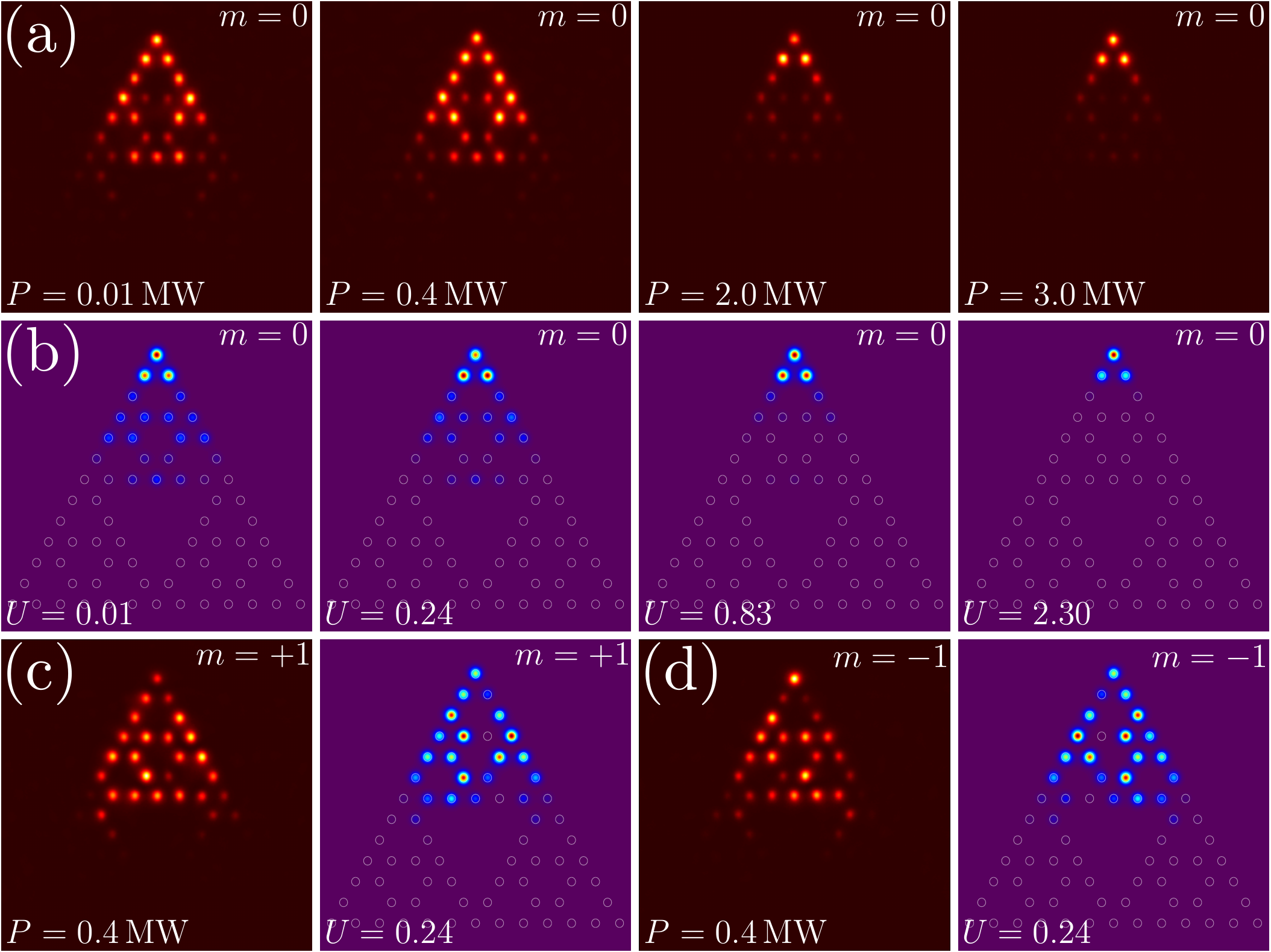}
	\caption{\textbf{Experimental excitations in topological trivial lattice.}
		(a) Output experimental intensity distributions (panels with maroon background) for trivial-phase excitation $(m=0)$ of three corner waveguides in fractal array with ${r=0.5a}$ at different input peak powers $P$ compared with (b) theoretical output distributions (panels with purple background). (c) and (d) show comparison of the experimental and theoretical output intensity distributions for vortex-carrying excitation of three corner waveguides with ${m=+1}$ (c) and ${m=-1}$ (d).}
	\label{fig3}
\end{figure}

To illustrate the importance of shift parameter $r$ for emergence of topological corner states, we first consider the array with ${r=0.5a}$, where all eigenmodes of the array are effectively extended. Using three-spot input with in-phase spots (that corresponds to ${m=0}$) focused into three corner waveguides reveals strong diffraction broadening at low peak powers ${P\sim0.01~\rm MW}$ (see Fig.~\ref{fig3}a showing corresponding experimental output intensity distributions with maroon background). Only with increase of input peak power to values ${P>2~\rm MW}$ one observes the contraction of light to three corner waveguides and formation of compact corner soliton that in this case occurs above considerable power threshold. Notice that for ${m=0}$ input the output pattern is symmetric with respect to ${x=0}$ axis. These observations are confirmed by theoretically calculated output intensity distributions at different dimensionless powers $U$ shown in Fig.~\ref{fig3}b (images with purple background). Using vortex-carrying inputs with ${m=+1}$ (Fig.~\ref{fig3}c) and ${m=-1}$ (Fig.~\ref{fig3}d) reveals stronger diffraction for the same peak low and moderate power levels $P$ that becomes notably asymmetric. The formation of corner vortex soliton occurs in this case above even higher power threshold ${P\sim 3~\rm MW}$. We observed similar outputs in arrays with ${r=0.4a}$.

\begin{figure}[h!]
	\centering
	\includegraphics[width=\columnwidth]{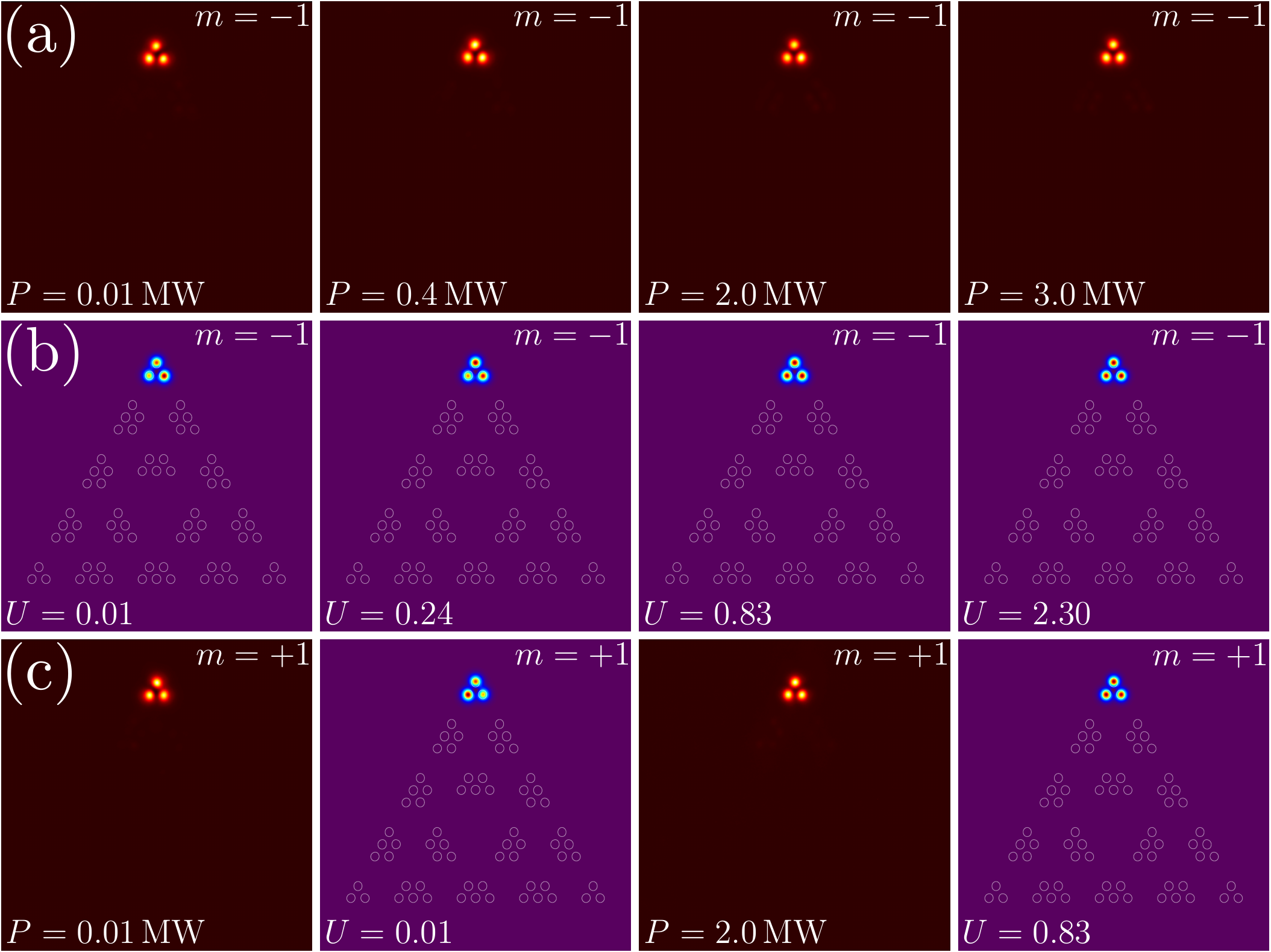}
	\caption{\textbf{Experimental excitation of corner vortex solitons.}
		Comparison of experimental (a) and theoretical (b) output intensity distributions for vortex-carrying excitation $(m=-1)$ of three corner waveguides in array with ${r=0.7a}$ at different peak powers. (c) The same for ${m=+1}$ excitation.}
	\label{fig4}
\end{figure}

Experimental evidence of thresholdless topological corner vortex solitons is presented in Fig.~\ref{fig4} in fractal array with shift parameter ${r=0.7a}$ that supports well-localized linear corner vortex states. When using three-spot input with ${m=-1}$ vortex, we observed (Fig.~\ref{fig4}a) that light remains confined predominantly in three corner waveguides across all power levels, from ${P=0.01~\rm MW}$ to $P=3~\rm MW$ (limited by the material damage threshold), and that vortical phase structure, with a singularity between three corner waveguides, is present in all cases. This confirms thresholdless and topological nature of such solitons (i.e. their bifurcation from in-gap linear vortex corner states), as well as their exceptional robustness. Notice that output intensity distributions are slightly distorted at low and moderate peak powers (i.e. they do not feature ideal $\mathcal{C}_3$ discrete rotational symmetry). This is a consequence of slight ellipticity of waveguide modes that cannot be eliminated completely even with ``multi-scan'' writing technique. Increasing peak power to ${P=2~\rm MW}$ practically eliminates this distortion. These observations are in agreement with theoretical simulations shown in Fig. \ref{fig4}b. Similar results for ${m=+1}$ vortex-carrying inputs (see Fig.~\ref{fig4}c) confirm the system's high symmetry.

As one can see from theoretical plots in Fig.~\ref{fig2}, coupling with extended states in the band should result in the appearance of long tails inside the array. We do not see such tails (that would clearly indicate on nonlinearity-induced reshaping) in experiments only because for chosen distortion parameter $r$ value they start to appear for peak powers ${P>3~\textrm{MW}}$ that is close to the material damage threshold. On this reason, to avoid sample damage we did not increase peak power well above ${3~\textrm{MW}}$. At the same time, by looking at distributions in Fig.~\ref{fig3} (in trivial case), one can see that nonlinear contribution to the refractive index is considerable at such peak powers and notably changes output intensity distributions, so this regime is indeed strongly nonlinear.

\begin{figure}[h!]
	\centering
	\includegraphics[width=\columnwidth]{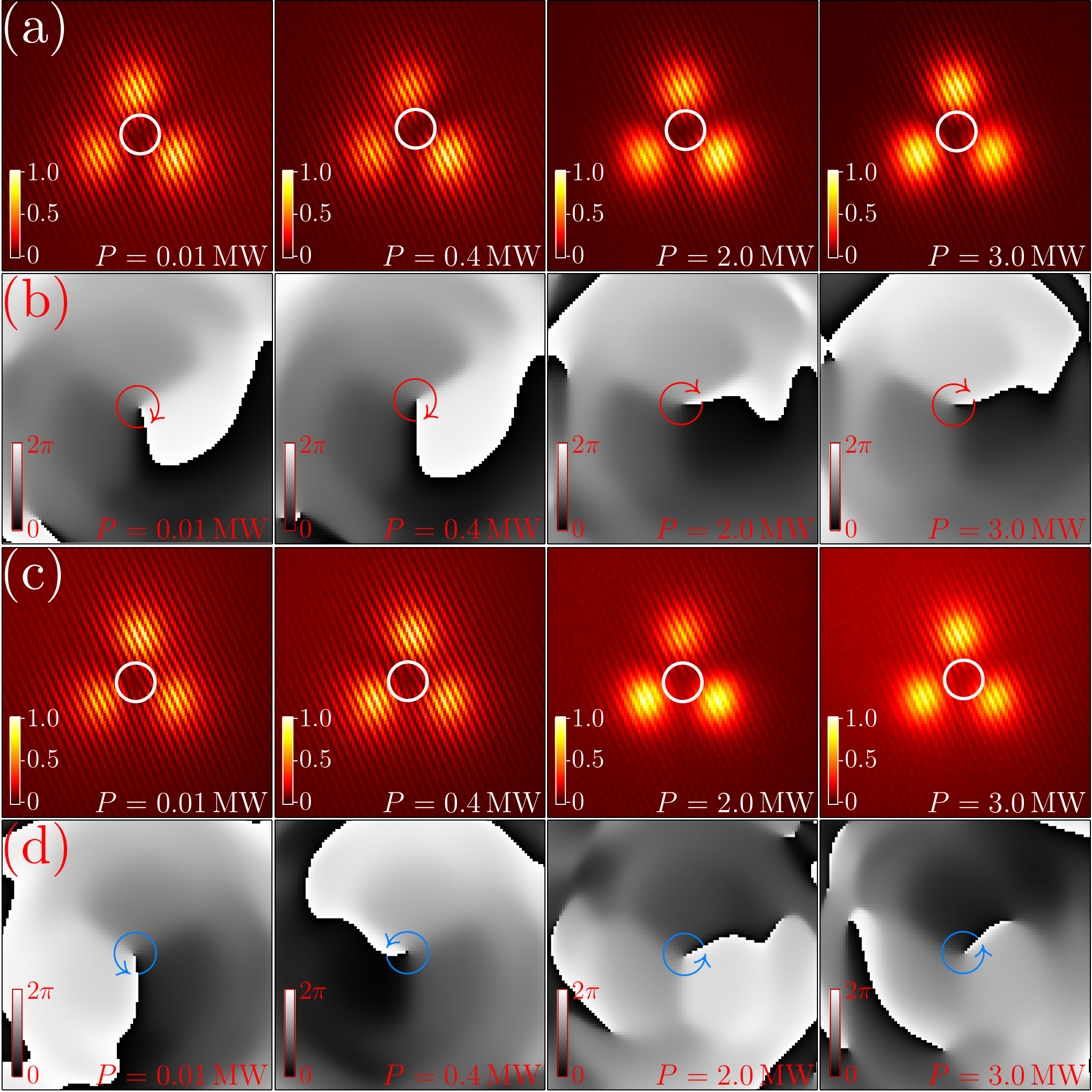}
	\caption{\textbf{Experimentally measured phase distribution of the excited corner vortex soliton.}
		(a,c) Experimentally measured single-shot interference patterns and corresponding retrieved phase distributions (b,d). Measured phases confirm the formation of corner vortex solitons with topological charge $m=-1$ (a,b) and $m=+1$ (c,d) at different peak powers $P$. White circles schematically denote the position of the phase singularity.}
	\label{fig5}
\end{figure}

To unequivocally confirm vortical structure of the output fields in Fig.~\ref{fig4}, we recorded and analyzed single-shot interference patterns of output fields with coherent plane wave for different peak powers to retrieve the output phase distributions (see Section S1 in the Supplementary Materials). Experimentally measured normalized fringe patterns and corresponding phase distributions, zoomed in around three corner waveguides, are shown in Figs.~\ref{fig5}a and \ref{fig5}b for ${m=-1}$ and in Figs.~\ref{fig5}c and \ref{fig5}d for ${m=+1}$ vortex solitons, respectively. The fork (denoted with white circle) in the interference pattern with orientation reflecting charge of the vortex is present between corner waveguides at all peak power levels. Retrieved phases conclusively confirm global vorticity imposed on the field for both ${m=-1}$ and ${m=+1}$ cases.

\section{Conclusions}
Summarizing, we have experimentally observed corner vortex solitons in nonlinear HOTIs based on fractal waveguide array and verified their vorticity by phase-resolved interferometry. The essential mechanism is the degeneracy-enabled construction of an in-gap corner vortex eigenstate, which yields thresholdless nonlinear continuation into robust vortex-soliton families. We found that observed vortex solitons in fractal arrays are exceptionally robust in a broad power range and that localization of the vortex-carrying modes can be changed via design of the underlying fractal structure. Our work opens a route to the investigation of diverse nonlinear phenomena in topologically nontrivial systems hosting vortices. This includes generation of new frequencies and states with higher topological charges, nonlinearity-controlled switching, and lasing.

These findings reveal a rich interplay among fractality, topology, orbital angular momentum degree of freedom, and nonlinearity. They open avenues for experimental studies of fractal topological systems in domains beyond photonics, such as acoustics and matter-wave physics. While obtained for self-focusing nonlinearity, our results can be generalized to the case of self-defocusing materials. Furthermore, instead of the waveguide array systems, topological fractal micropillar arrays could be realized in exciton-polariton microcavities or other dissipative/cavity platforms~\cite{klembt.nature.562.552.2018}. Under resonant pumping~\cite{kartashov.prl.119.253904.2017}, such systems may support lasing in corner vortex solitons and exhibit bistability effects with vortices. The system proposed here could be generalized to nanoscale settings, to achieve the formation of topologically protected nano-vortices, by analogy with recent experiments with lasing on topological disclinations~\cite{hwang.np.18.286.2024}. Finally, vortex-carrying states and new aspects of their localization associated with fractality of topological structure could be explored in $p$-orbital HOTIs \cite{zhang.elight.3.5.2023, bongiovanni.lpr.18.2400638.2024}.

\begin{acknowledgments}		
This work was supported by the Natural Science Basic Research Program of Shaanxi Province (2024JC-JCQN-06, 2025JC-QYCX-006), the National Natural Science Foundation of China (12474337), the Russian Science Foundation (grant 24-12-00167) and partially by the project FFUU-2024-0003 of the Institute of Spectroscopy of RAS.
\end{acknowledgments}

%\bibliographystyle{myprl}
%\bibliography{my_library1,my_library2}

\begin{thebibliography}{62}%
	\makeatletter
	\providecommand \@ifxundefined [1]{%
		\@ifx{#1\undefined}
	}%
	\providecommand \@ifnum [1]{%
		\ifnum #1\expandafter \@firstoftwo
		\else \expandafter \@secondoftwo
		\fi
	}%
	\providecommand \@ifx [1]{%
		\ifx #1\expandafter \@firstoftwo
		\else \expandafter \@secondoftwo
		\fi
	}%
	\providecommand \natexlab [1]{#1}%
	\providecommand \enquote  [1]{``#1''}%
	\providecommand \bibnamefont  [1]{#1}%
	\providecommand \bibfnamefont [1]{#1}%
	\providecommand \citenamefont [1]{#1}%
	\providecommand \href@noop [0]{\@secondoftwo}%
	\providecommand \href [0]{\begingroup \@sanitize@url \@href}%
	\providecommand \@href[1]{\@@startlink{#1}\@@href}%
	\providecommand \@@href[1]{\endgroup#1\@@endlink}%
	\providecommand \@sanitize@url [0]{\catcode `\\12\catcode `\$12\catcode
		`\&12\catcode `\#12\catcode `\^12\catcode `\_12\catcode `\%12\relax}%
	\providecommand \@@startlink[1]{}%
	\providecommand \@@endlink[0]{}%
	\providecommand \url  [0]{\begingroup\@sanitize@url \@url }%
	\providecommand \@url [1]{\endgroup\@href {#1}{\urlprefix }}%
	\providecommand \urlprefix  [0]{URL }%
	\providecommand \Eprint [0]{\href }%
	\providecommand \doibase [0]{https://doi.org/}%
	\providecommand \selectlanguage [0]{\@gobble}%
	\providecommand \bibinfo  [0]{\@secondoftwo}%
	\providecommand \bibfield  [0]{\@secondoftwo}%
	\providecommand \translation [1]{[#1]}%
	\providecommand \BibitemOpen [0]{}%
	\providecommand \bibitemStop [0]{}%
	\providecommand \bibitemNoStop [0]{.\EOS\space}%
	\providecommand \EOS [0]{\spacefactor3000\relax}%
	\providecommand \BibitemShut  [1]{\csname bibitem#1\endcsname}%
	\let\auto@bib@innerbib\@empty
	%</preamble>
	\bibitem [{\citenamefont {Lu}\ \emph {et~al.}(2014)\citenamefont {Lu},
		\citenamefont {Joannopoulos},\ and\ \citenamefont
		{Solja{\v{c}}i{\'c}}}]{lu.np.8.821.2014}%
	\BibitemOpen
	\bibfield  {author} {\bibinfo {author} {\bibfnamefont {L.}~\bibnamefont
			{Lu}}, \bibinfo {author} {\bibfnamefont {J.~D.}\ \bibnamefont
			{Joannopoulos}},\ and\ \bibinfo {author} {\bibfnamefont {M.}~\bibnamefont
			{Solja{\v{c}}i{\'c}}},\ }\bibfield  {title} {\bibinfo {title} {Topological
			photonics},\ }\href {https://doi.org/10.1038/nphoton.2014.248} {\bibfield
		{journal} {\bibinfo  {journal} {Nat. Photon.}\ }\textbf {\bibinfo {volume}
			{8}},\ \bibinfo {pages} {821} (\bibinfo {year} {2014})}\BibitemShut {NoStop}%
	\bibitem [{\citenamefont {Ozawa}\ \emph {et~al.}(2019)\citenamefont {Ozawa},
		\citenamefont {Price}, \citenamefont {Amo}, \citenamefont {Goldman},
		\citenamefont {Hafezi}, \citenamefont {Lu}, \citenamefont {Rechtsman},
		\citenamefont {Schuster}, \citenamefont {Simon}, \citenamefont {Zilberberg},\
		and\ \citenamefont {Carusotto}}]{ozawa.rmp.91.015006.2019}%
	\BibitemOpen
	\bibfield  {author} {\bibinfo {author} {\bibfnamefont {T.}~\bibnamefont
			{Ozawa}}, \bibinfo {author} {\bibfnamefont {H.~M.}\ \bibnamefont {Price}},
		\bibinfo {author} {\bibfnamefont {A.}~\bibnamefont {Amo}}, \bibinfo {author}
		{\bibfnamefont {N.}~\bibnamefont {Goldman}}, \bibinfo {author} {\bibfnamefont
			{M.}~\bibnamefont {Hafezi}}, \bibinfo {author} {\bibfnamefont
			{L.}~\bibnamefont {Lu}}, \bibinfo {author} {\bibfnamefont {M.~C.}\
			\bibnamefont {Rechtsman}}, \bibinfo {author} {\bibfnamefont {D.}~\bibnamefont
			{Schuster}}, \bibinfo {author} {\bibfnamefont {J.}~\bibnamefont {Simon}},
		\bibinfo {author} {\bibfnamefont {O.}~\bibnamefont {Zilberberg}},\ and\
		\bibinfo {author} {\bibfnamefont {I.}~\bibnamefont {Carusotto}},\ }\bibfield
	{title} {\bibinfo {title} {Topological photonics},\ }\href
	{https://doi.org/10.1103/RevModPhys.91.015006} {\bibfield  {journal}
		{\bibinfo  {journal} {Rev. Mod. Phys.}\ }\textbf {\bibinfo {volume} {91}},\
		\bibinfo {pages} {015006} (\bibinfo {year} {2019})}\BibitemShut {NoStop}%
	\bibitem [{\citenamefont {Zhang}\ \emph
		{et~al.}(2023{\natexlab{a}})\citenamefont {Zhang}, \citenamefont
		{Zangeneh-Nejad}, \citenamefont {Chen}, \citenamefont {Lu},\ and\
		\citenamefont {Christensen}}]{zhang.nature.618.687.2023}%
	\BibitemOpen
	\bibfield  {author} {\bibinfo {author} {\bibfnamefont {X.}~\bibnamefont
			{Zhang}}, \bibinfo {author} {\bibfnamefont {F.}~\bibnamefont
			{Zangeneh-Nejad}}, \bibinfo {author} {\bibfnamefont {Z.-G.}\ \bibnamefont
			{Chen}}, \bibinfo {author} {\bibfnamefont {M.-H.}\ \bibnamefont {Lu}},\ and\
		\bibinfo {author} {\bibfnamefont {J.}~\bibnamefont {Christensen}},\
	}\bibfield  {title} {\bibinfo {title} {A second wave of topological phenomena
			in photonics and acoustics},\ }\href
	{https://doi.org/10.1038/s41586-023-06163-9} {\bibfield  {journal} {\bibinfo
			{journal} {Nature}\ }\textbf {\bibinfo {volume} {618}},\ \bibinfo {pages}
		{687} (\bibinfo {year} {2023}{\natexlab{a}})}\BibitemShut {NoStop}%
	\bibitem [{\citenamefont {Leykam}\ \emph {et~al.}(2026)\citenamefont {Leykam},
		\citenamefont {Xue}, \citenamefont {Zhang},\ and\ \citenamefont
		{Chong}}]{leykam.nrp.8.55.2026}%
	\BibitemOpen
	\bibfield  {author} {\bibinfo {author} {\bibfnamefont {D.}~\bibnamefont
			{Leykam}}, \bibinfo {author} {\bibfnamefont {H.}~\bibnamefont {Xue}},
		\bibinfo {author} {\bibfnamefont {B.}~\bibnamefont {Zhang}},\ and\ \bibinfo
		{author} {\bibfnamefont {Y.~D.}\ \bibnamefont {Chong}},\ }\bibfield  {title}
	{\bibinfo {title} {Limitations and possibilities of topological photonics},\
	}\href {https://doi.org/10.1038/s42254-025-00889-3} {\bibfield  {journal}
		{\bibinfo  {journal} {Nat. Rev. Phys.}\ }\textbf {\bibinfo {volume} {8}},\
		\bibinfo {pages} {55} (\bibinfo {year} {2026})}\BibitemShut {NoStop}%
	\bibitem [{\citenamefont {Wieder}\ \emph {et~al.}(2022)\citenamefont {Wieder},
		\citenamefont {Bradlyn}, \citenamefont {Cano}, \citenamefont {Wang},
		\citenamefont {Vergniory}, \citenamefont {Elcoro}, \citenamefont {Soluyanov},
		\citenamefont {Felser}, \citenamefont {Neupert}, \citenamefont {Regnault},\
		and\ \citenamefont {Bernevig}}]{wieder.nrm.7.196.2022}%
	\BibitemOpen
	\bibfield  {author} {\bibinfo {author} {\bibfnamefont {B.~J.}\ \bibnamefont
			{Wieder}}, \bibinfo {author} {\bibfnamefont {B.}~\bibnamefont {Bradlyn}},
		\bibinfo {author} {\bibfnamefont {J.}~\bibnamefont {Cano}}, \bibinfo {author}
		{\bibfnamefont {Z.}~\bibnamefont {Wang}}, \bibinfo {author} {\bibfnamefont
			{M.~G.}\ \bibnamefont {Vergniory}}, \bibinfo {author} {\bibfnamefont
			{L.}~\bibnamefont {Elcoro}}, \bibinfo {author} {\bibfnamefont {A.~A.}\
			\bibnamefont {Soluyanov}}, \bibinfo {author} {\bibfnamefont {C.}~\bibnamefont
			{Felser}}, \bibinfo {author} {\bibfnamefont {T.}~\bibnamefont {Neupert}},
		\bibinfo {author} {\bibfnamefont {N.}~\bibnamefont {Regnault}},\ and\
		\bibinfo {author} {\bibfnamefont {B.~A.}\ \bibnamefont {Bernevig}},\
	}\bibfield  {title} {\bibinfo {title} {Topological materials discovery from
			crystal symmetry},\ }\href {https://doi.org/10.1038/s41578-021-00380-2}
	{\bibfield  {journal} {\bibinfo  {journal} {Nat. Rev. Mater.}\ }\textbf
		{\bibinfo {volume} {7}},\ \bibinfo {pages} {196} (\bibinfo {year}
		{2022})}\BibitemShut {NoStop}%
	\bibitem [{\citenamefont {Xie}\ \emph {et~al.}(2021)\citenamefont {Xie},
		\citenamefont {Wang}, \citenamefont {Zhang}, \citenamefont {Zhan},
		\citenamefont {Jiang}, \citenamefont {Lu},\ and\ \citenamefont
		{Chen}}]{xie.nrp.3.520.2021}%
	\BibitemOpen
	\bibfield  {author} {\bibinfo {author} {\bibfnamefont {B.}~\bibnamefont
			{Xie}}, \bibinfo {author} {\bibfnamefont {H.-X.}\ \bibnamefont {Wang}},
		\bibinfo {author} {\bibfnamefont {X.}~\bibnamefont {Zhang}}, \bibinfo
		{author} {\bibfnamefont {P.}~\bibnamefont {Zhan}}, \bibinfo {author}
		{\bibfnamefont {J.-H.}\ \bibnamefont {Jiang}}, \bibinfo {author}
		{\bibfnamefont {M.}~\bibnamefont {Lu}},\ and\ \bibinfo {author}
		{\bibfnamefont {Y.}~\bibnamefont {Chen}},\ }\bibfield  {title} {\bibinfo
		{title} {Higher-order band topology},\ }\href
	{https://doi.org/10.1038/s42254-021-00323-4} {\bibfield  {journal} {\bibinfo
			{journal} {Nat. Rev. Phys.}\ }\textbf {\bibinfo {volume} {3}},\ \bibinfo
		{pages} {520} (\bibinfo {year} {2021})}\BibitemShut {NoStop}%
	\bibitem [{\citenamefont {Lin}\ \emph {et~al.}(2023)\citenamefont {Lin},
		\citenamefont {Wang}, \citenamefont {Liu}, \citenamefont {Xue}, \citenamefont
		{Zhang}, \citenamefont {Chong},\ and\ \citenamefont
		{Jiang}}]{lin.nrp.5.483.2023}%
	\BibitemOpen
	\bibfield  {author} {\bibinfo {author} {\bibfnamefont {Z.-K.}\ \bibnamefont
			{Lin}}, \bibinfo {author} {\bibfnamefont {Q.}~\bibnamefont {Wang}}, \bibinfo
		{author} {\bibfnamefont {Y.}~\bibnamefont {Liu}}, \bibinfo {author}
		{\bibfnamefont {H.}~\bibnamefont {Xue}}, \bibinfo {author} {\bibfnamefont
			{B.}~\bibnamefont {Zhang}}, \bibinfo {author} {\bibfnamefont
			{Y.}~\bibnamefont {Chong}},\ and\ \bibinfo {author} {\bibfnamefont {J.-H.}\
			\bibnamefont {Jiang}},\ }\bibfield  {title} {\bibinfo {title} {Topological
			phenomena at defects in acoustic, photonic and solid-state lattices},\ }\href
	{https://doi.org/10.1038/s42254-023-00602-2} {\bibfield  {journal} {\bibinfo
			{journal} {Nat. Rev. Phys.}\ }\textbf {\bibinfo {volume} {5}},\ \bibinfo
		{pages} {483} (\bibinfo {year} {2023})}\BibitemShut {NoStop}%
	\bibitem [{\citenamefont {Benalcazar}\ \emph {et~al.}(2017)\citenamefont
		{Benalcazar}, \citenamefont {Bernevig},\ and\ \citenamefont
		{Hughes}}]{banalcazar.science.357.61.2017}%
	\BibitemOpen
	\bibfield  {author} {\bibinfo {author} {\bibfnamefont {W.~A.}\ \bibnamefont
			{Benalcazar}}, \bibinfo {author} {\bibfnamefont {B.~A.}\ \bibnamefont
			{Bernevig}},\ and\ \bibinfo {author} {\bibfnamefont {T.~L.}\ \bibnamefont
			{Hughes}},\ }\bibfield  {title} {\bibinfo {title} {Quantized electric
			multipole insulators},\ }\href {https://doi.org/10.1126/science.aah6442}
	{\bibfield  {journal} {\bibinfo  {journal} {Science}\ }\textbf {\bibinfo
			{volume} {357}},\ \bibinfo {pages} {61} (\bibinfo {year} {2017})}\BibitemShut
	{NoStop}%
	\bibitem [{\citenamefont {Serra-Garcia}\ \emph {et~al.}(2018)\citenamefont
		{Serra-Garcia}, \citenamefont {Peri}, \citenamefont {S{\"u}sstrunk},
		\citenamefont {Bilal}, \citenamefont {Larsen}, \citenamefont {Villanueva},\
		and\ \citenamefont {Huber}}]{serra.nature.555.342.2018}%
	\BibitemOpen
	\bibfield  {author} {\bibinfo {author} {\bibfnamefont {M.}~\bibnamefont
			{Serra-Garcia}}, \bibinfo {author} {\bibfnamefont {V.}~\bibnamefont {Peri}},
		\bibinfo {author} {\bibfnamefont {R.}~\bibnamefont {S{\"u}sstrunk}}, \bibinfo
		{author} {\bibfnamefont {O.~R.}\ \bibnamefont {Bilal}}, \bibinfo {author}
		{\bibfnamefont {T.}~\bibnamefont {Larsen}}, \bibinfo {author} {\bibfnamefont
			{L.~G.}\ \bibnamefont {Villanueva}},\ and\ \bibinfo {author} {\bibfnamefont
			{S.~D.}\ \bibnamefont {Huber}},\ }\bibfield  {title} {\bibinfo {title}
		{Observation of a phononic quadrupole topological insulator},\ }\href
	{https://doi.org/10.1038/nature25156} {\bibfield  {journal} {\bibinfo
			{journal} {Nature}\ }\textbf {\bibinfo {volume} {555}},\ \bibinfo {pages}
		{342} (\bibinfo {year} {2018})}\BibitemShut {NoStop}%
	\bibitem [{\citenamefont {Noh}\ \emph {et~al.}(2018)\citenamefont {Noh},
		\citenamefont {Benalcazar}, \citenamefont {Huang}, \citenamefont {Collins},
		\citenamefont {Chen}, \citenamefont {Hughes},\ and\ \citenamefont
		{Rechtsman}}]{noh.np.12.408.2018}%
	\BibitemOpen
	\bibfield  {author} {\bibinfo {author} {\bibfnamefont {J.}~\bibnamefont
			{Noh}}, \bibinfo {author} {\bibfnamefont {W.~A.}\ \bibnamefont {Benalcazar}},
		\bibinfo {author} {\bibfnamefont {S.}~\bibnamefont {Huang}}, \bibinfo
		{author} {\bibfnamefont {M.~J.}\ \bibnamefont {Collins}}, \bibinfo {author}
		{\bibfnamefont {K.~P.}\ \bibnamefont {Chen}}, \bibinfo {author}
		{\bibfnamefont {T.~L.}\ \bibnamefont {Hughes}},\ and\ \bibinfo {author}
		{\bibfnamefont {M.~C.}\ \bibnamefont {Rechtsman}},\ }\bibfield  {title}
	{\bibinfo {title} {Topological protection of photonic mid-gap defect modes},\
	}\href {https://doi.org/10.1038/s41566-018-0179-3} {\bibfield  {journal}
		{\bibinfo  {journal} {Nat. Photon.}\ }\textbf {\bibinfo {volume} {12}},\
		\bibinfo {pages} {408} (\bibinfo {year} {2018})}\BibitemShut {NoStop}%
	\bibitem [{\citenamefont {Xie}\ \emph {et~al.}(2019)\citenamefont {Xie},
		\citenamefont {Su}, \citenamefont {Wang}, \citenamefont {Su}, \citenamefont
		{Shen}, \citenamefont {Zhan}, \citenamefont {Lu}, \citenamefont {Wang},\ and\
		\citenamefont {Chen}}]{xie.prl.122.233903.2019}%
	\BibitemOpen
	\bibfield  {author} {\bibinfo {author} {\bibfnamefont {B.-Y.}\ \bibnamefont
			{Xie}}, \bibinfo {author} {\bibfnamefont {G.-X.}\ \bibnamefont {Su}},
		\bibinfo {author} {\bibfnamefont {H.-F.}\ \bibnamefont {Wang}}, \bibinfo
		{author} {\bibfnamefont {H.}~\bibnamefont {Su}}, \bibinfo {author}
		{\bibfnamefont {X.-P.}\ \bibnamefont {Shen}}, \bibinfo {author}
		{\bibfnamefont {P.}~\bibnamefont {Zhan}}, \bibinfo {author} {\bibfnamefont
			{M.-H.}\ \bibnamefont {Lu}}, \bibinfo {author} {\bibfnamefont {Z.-L.}\
			\bibnamefont {Wang}},\ and\ \bibinfo {author} {\bibfnamefont {Y.-F.}\
			\bibnamefont {Chen}},\ }\bibfield  {title} {\bibinfo {title} {Visualization
			of higher-order topological insulating phases in two-dimensional dielectric
			photonic crystals},\ }\href {https://doi.org/10.1103/PhysRevLett.122.233903}
	{\bibfield  {journal} {\bibinfo  {journal} {Phys. Rev. Lett.}\ }\textbf
		{\bibinfo {volume} {122}},\ \bibinfo {pages} {233903} (\bibinfo {year}
		{2019})}\BibitemShut {NoStop}%
	\bibitem [{\citenamefont {Chen}\ \emph {et~al.}(2019)\citenamefont {Chen},
		\citenamefont {Deng}, \citenamefont {Shi}, \citenamefont {Zhao},
		\citenamefont {Chen},\ and\ \citenamefont {Dong}}]{chen.prl.122.233902.2019}%
	\BibitemOpen
	\bibfield  {author} {\bibinfo {author} {\bibfnamefont {X.-D.}\ \bibnamefont
			{Chen}}, \bibinfo {author} {\bibfnamefont {W.-M.}\ \bibnamefont {Deng}},
		\bibinfo {author} {\bibfnamefont {F.-L.}\ \bibnamefont {Shi}}, \bibinfo
		{author} {\bibfnamefont {F.-L.}\ \bibnamefont {Zhao}}, \bibinfo {author}
		{\bibfnamefont {M.}~\bibnamefont {Chen}},\ and\ \bibinfo {author}
		{\bibfnamefont {J.-W.}\ \bibnamefont {Dong}},\ }\bibfield  {title} {\bibinfo
		{title} {Direct observation of corner states in second-order topological
			photonic crystal slabs},\ }\href
	{https://doi.org/10.1103/PhysRevLett.122.233902} {\bibfield  {journal}
		{\bibinfo  {journal} {Phys. Rev. Lett.}\ }\textbf {\bibinfo {volume} {122}},\
		\bibinfo {pages} {233902} (\bibinfo {year} {2019})}\BibitemShut {NoStop}%
	\bibitem [{\citenamefont {El~Hassan}\ \emph {et~al.}(2019)\citenamefont
		{El~Hassan}, \citenamefont {Kunst}, \citenamefont {Moritz}, \citenamefont
		{Andler}, \citenamefont {Bergholtz},\ and\ \citenamefont
		{Bourennane}}]{hassan.np.13.697.2019}%
	\BibitemOpen
	\bibfield  {author} {\bibinfo {author} {\bibfnamefont {A.}~\bibnamefont
			{El~Hassan}}, \bibinfo {author} {\bibfnamefont {F.~K.}\ \bibnamefont
			{Kunst}}, \bibinfo {author} {\bibfnamefont {A.}~\bibnamefont {Moritz}},
		\bibinfo {author} {\bibfnamefont {G.}~\bibnamefont {Andler}}, \bibinfo
		{author} {\bibfnamefont {E.~J.}\ \bibnamefont {Bergholtz}},\ and\ \bibinfo
		{author} {\bibfnamefont {M.}~\bibnamefont {Bourennane}},\ }\bibfield  {title}
	{\bibinfo {title} {Corner states of light in photonic waveguides},\ }\href
	{https://doi.org/10.1038/s41566-019-0519-y} {\bibfield  {journal} {\bibinfo
			{journal} {Nat. Photon.}\ }\textbf {\bibinfo {volume} {13}},\ \bibinfo
		{pages} {697} (\bibinfo {year} {2019})}\BibitemShut {NoStop}%
	\bibitem [{\citenamefont {Li}\ \emph {et~al.}(2020{\natexlab{a}})\citenamefont
		{Li}, \citenamefont {Zhirihin}, \citenamefont {Gorlach}, \citenamefont {Ni},
		\citenamefont {Filonov}, \citenamefont {Slobozhanyuk}, \citenamefont
		{Al\`u},\ and\ \citenamefont {Khanikaev}}]{li.np.14.89.2020}%
	\BibitemOpen
	\bibfield  {author} {\bibinfo {author} {\bibfnamefont {M.}~\bibnamefont
			{Li}}, \bibinfo {author} {\bibfnamefont {D.}~\bibnamefont {Zhirihin}},
		\bibinfo {author} {\bibfnamefont {M.}~\bibnamefont {Gorlach}}, \bibinfo
		{author} {\bibfnamefont {X.}~\bibnamefont {Ni}}, \bibinfo {author}
		{\bibfnamefont {D.}~\bibnamefont {Filonov}}, \bibinfo {author} {\bibfnamefont
			{A.}~\bibnamefont {Slobozhanyuk}}, \bibinfo {author} {\bibfnamefont
			{A.}~\bibnamefont {Al\`u}},\ and\ \bibinfo {author} {\bibfnamefont {A.~B.}\
			\bibnamefont {Khanikaev}},\ }\bibfield  {title} {\bibinfo {title}
		{Higher-order topological states in photonic kagome crystals with long-range
			interactions},\ }\href {https://doi.org/10.1038/s41566-019-0561-9} {\bibfield
		{journal} {\bibinfo  {journal} {Nat. Photon.}\ }\textbf {\bibinfo {volume}
			{14}},\ \bibinfo {pages} {89} (\bibinfo {year}
		{2020}{\natexlab{a}})}\BibitemShut {NoStop}%
	\bibitem [{\citenamefont {Benalcazar}\ and\ \citenamefont
		{Cerjan}(2022)}]{benalcazar.prl.128.127601.2022}%
	\BibitemOpen
	\bibfield  {author} {\bibinfo {author} {\bibfnamefont {W.~A.}\ \bibnamefont
			{Benalcazar}}\ and\ \bibinfo {author} {\bibfnamefont {A.}~\bibnamefont
			{Cerjan}},\ }\bibfield  {title} {\bibinfo {title} {Chiral-symmetric
			higher-order topological phases of matter},\ }\href
	{https://doi.org/10.1103/PhysRevLett.128.127601} {\bibfield  {journal}
		{\bibinfo  {journal} {Phys. Rev. Lett.}\ }\textbf {\bibinfo {volume} {128}},\
		\bibinfo {pages} {127601} (\bibinfo {year} {2022})}\BibitemShut {NoStop}%
	\bibitem [{\citenamefont {Peterson}\ \emph {et~al.}(2018)\citenamefont
		{Peterson}, \citenamefont {Benalcazar}, \citenamefont {Hughes},\ and\
		\citenamefont {Bahl}}]{peterson.nature.555.346.2018}%
	\BibitemOpen
	\bibfield  {author} {\bibinfo {author} {\bibfnamefont {C.~W.}\ \bibnamefont
			{Peterson}}, \bibinfo {author} {\bibfnamefont {W.~A.}\ \bibnamefont
			{Benalcazar}}, \bibinfo {author} {\bibfnamefont {T.~L.}\ \bibnamefont
			{Hughes}},\ and\ \bibinfo {author} {\bibfnamefont {G.}~\bibnamefont {Bahl}},\
	}\bibfield  {title} {\bibinfo {title} {A quantized microwave quadrupole
			insulator with topologically protected corner states},\ }\href
	{https://doi.org/10.1038/nature25777} {\bibfield  {journal} {\bibinfo
			{journal} {Nature}\ }\textbf {\bibinfo {volume} {555}},\ \bibinfo {pages}
		{346} (\bibinfo {year} {2018})}\BibitemShut {NoStop}%
	\bibitem [{\citenamefont {Mittal}\ \emph {et~al.}(2019)\citenamefont {Mittal},
		\citenamefont {Orre}, \citenamefont {Zhu}, \citenamefont {Gorlach},
		\citenamefont {Poddubny},\ and\ \citenamefont
		{Hafezi}}]{mittal.np.13.692.2019}%
	\BibitemOpen
	\bibfield  {author} {\bibinfo {author} {\bibfnamefont {S.}~\bibnamefont
			{Mittal}}, \bibinfo {author} {\bibfnamefont {V.~V.}\ \bibnamefont {Orre}},
		\bibinfo {author} {\bibfnamefont {G.}~\bibnamefont {Zhu}}, \bibinfo {author}
		{\bibfnamefont {M.~A.}\ \bibnamefont {Gorlach}}, \bibinfo {author}
		{\bibfnamefont {A.}~\bibnamefont {Poddubny}},\ and\ \bibinfo {author}
		{\bibfnamefont {M.}~\bibnamefont {Hafezi}},\ }\bibfield  {title} {\bibinfo
		{title} {Photonic quadrupole topological phases},\ }\href
	{https://doi.org/10.1038/s41566-019-0452-0} {\bibfield  {journal} {\bibinfo
			{journal} {Nat. Photon.}\ }\textbf {\bibinfo {volume} {13}},\ \bibinfo
		{pages} {692} (\bibinfo {year} {2019})}\BibitemShut {NoStop}%
	\bibitem [{\citenamefont {Zhang}\ \emph {et~al.}(2019)\citenamefont {Zhang},
		\citenamefont {Wang}, \citenamefont {Lin}, \citenamefont {Tian},
		\citenamefont {Xie}, \citenamefont {Lu}, \citenamefont {Chen},\ and\
		\citenamefont {Jiang}}]{zhang.np.15.582.2019}%
	\BibitemOpen
	\bibfield  {author} {\bibinfo {author} {\bibfnamefont {X.}~\bibnamefont
			{Zhang}}, \bibinfo {author} {\bibfnamefont {H.-X.}\ \bibnamefont {Wang}},
		\bibinfo {author} {\bibfnamefont {Z.-K.}\ \bibnamefont {Lin}}, \bibinfo
		{author} {\bibfnamefont {Y.}~\bibnamefont {Tian}}, \bibinfo {author}
		{\bibfnamefont {B.}~\bibnamefont {Xie}}, \bibinfo {author} {\bibfnamefont
			{M.-H.}\ \bibnamefont {Lu}}, \bibinfo {author} {\bibfnamefont {Y.-F.}\
			\bibnamefont {Chen}},\ and\ \bibinfo {author} {\bibfnamefont {J.-H.}\
			\bibnamefont {Jiang}},\ }\bibfield  {title} {\bibinfo {title} {Second-order
			topology and multidimensional topological transitions in sonic crystals},\
	}\href {https://doi.org/10.1038/s41567-019-0472-1} {\bibfield  {journal}
		{\bibinfo  {journal} {Nat. Phys.}\ }\textbf {\bibinfo {volume} {15}},\
		\bibinfo {pages} {582} (\bibinfo {year} {2019})}\BibitemShut {NoStop}%
	\bibitem [{\citenamefont {Peterson}\ \emph {et~al.}(2020)\citenamefont
		{Peterson}, \citenamefont {Li}, \citenamefont {Benalcazar}, \citenamefont
		{Hughes},\ and\ \citenamefont {Bahl}}]{peterson.science.368.1114.2020}%
	\BibitemOpen
	\bibfield  {author} {\bibinfo {author} {\bibfnamefont {C.~W.}\ \bibnamefont
			{Peterson}}, \bibinfo {author} {\bibfnamefont {T.}~\bibnamefont {Li}},
		\bibinfo {author} {\bibfnamefont {W.~A.}\ \bibnamefont {Benalcazar}},
		\bibinfo {author} {\bibfnamefont {T.~L.}\ \bibnamefont {Hughes}},\ and\
		\bibinfo {author} {\bibfnamefont {G.}~\bibnamefont {Bahl}},\ }\bibfield
	{title} {\bibinfo {title} {A fractional corner anomaly reveals higher-order
			topology},\ }\href {https://doi.org/10.1126/science.aba7604} {\bibfield
		{journal} {\bibinfo  {journal} {Science}\ }\textbf {\bibinfo {volume}
			{368}},\ \bibinfo {pages} {1114} (\bibinfo {year} {2020})}\BibitemShut
	{NoStop}%
	\bibitem [{\citenamefont {Peterson}\ \emph {et~al.}(2021)\citenamefont
		{Peterson}, \citenamefont {Li}, \citenamefont {Jiang}, \citenamefont
		{Hughes},\ and\ \citenamefont {Bahl}}]{peterson.nature.589.376.2021}%
	\BibitemOpen
	\bibfield  {author} {\bibinfo {author} {\bibfnamefont {C.~W.}\ \bibnamefont
			{Peterson}}, \bibinfo {author} {\bibfnamefont {T.}~\bibnamefont {Li}},
		\bibinfo {author} {\bibfnamefont {W.}~\bibnamefont {Jiang}}, \bibinfo
		{author} {\bibfnamefont {T.~L.}\ \bibnamefont {Hughes}},\ and\ \bibinfo
		{author} {\bibfnamefont {G.}~\bibnamefont {Bahl}},\ }\bibfield  {title}
	{\bibinfo {title} {Trapped fractional charges at bulk defects in topological
			insulators},\ }\href {https://doi.org/10.1038/s41586-020-03117-3} {\bibfield
		{journal} {\bibinfo  {journal} {Nature}\ }\textbf {\bibinfo {volume} {589}},\
		\bibinfo {pages} {376} (\bibinfo {year} {2021})}\BibitemShut {NoStop}%
	\bibitem [{\citenamefont {Liu}\ \emph {et~al.}(2021)\citenamefont {Liu},
		\citenamefont {Leung}, \citenamefont {Li}, \citenamefont {Lin}, \citenamefont
		{Tao}, \citenamefont {Poo},\ and\ \citenamefont
		{Jiang}}]{liu.nature.589.381.2021}%
	\BibitemOpen
	\bibfield  {author} {\bibinfo {author} {\bibfnamefont {Y.}~\bibnamefont
			{Liu}}, \bibinfo {author} {\bibfnamefont {S.}~\bibnamefont {Leung}}, \bibinfo
		{author} {\bibfnamefont {F.-F.}\ \bibnamefont {Li}}, \bibinfo {author}
		{\bibfnamefont {Z.-K.}\ \bibnamefont {Lin}}, \bibinfo {author} {\bibfnamefont
			{X.}~\bibnamefont {Tao}}, \bibinfo {author} {\bibfnamefont {Y.}~\bibnamefont
			{Poo}},\ and\ \bibinfo {author} {\bibfnamefont {J.-H.}\ \bibnamefont
			{Jiang}},\ }\bibfield  {title} {\bibinfo {title} {Bulk-disclination
			correspondence in topological crystalline insulators},\ }\href
	{https://doi.org/10.1038/s41586-020-03125-3} {\bibfield  {journal} {\bibinfo
			{journal} {Nature}\ }\textbf {\bibinfo {volume} {589}},\ \bibinfo {pages}
		{381} (\bibinfo {year} {2021})}\BibitemShut {NoStop}%
	\bibitem [{\citenamefont {Li}\ \emph {et~al.}(2020{\natexlab{b}})\citenamefont
		{Li}, \citenamefont {Zhu}, \citenamefont {Benalcazar},\ and\ \citenamefont
		{Hughes}}]{li.prb.101.115115.2020}%
	\BibitemOpen
	\bibfield  {author} {\bibinfo {author} {\bibfnamefont {T.}~\bibnamefont
			{Li}}, \bibinfo {author} {\bibfnamefont {P.}~\bibnamefont {Zhu}}, \bibinfo
		{author} {\bibfnamefont {W.~A.}\ \bibnamefont {Benalcazar}},\ and\ \bibinfo
		{author} {\bibfnamefont {T.~L.}\ \bibnamefont {Hughes}},\ }\bibfield  {title}
	{\bibinfo {title} {Fractional disclination charge in two-dimensional
			${C}_{n}$-symmetric topological crystalline insulators},\ }\href
	{https://doi.org/10.1103/PhysRevB.101.115115} {\bibfield  {journal} {\bibinfo
			{journal} {Phys. Rev. B}\ }\textbf {\bibinfo {volume} {101}},\ \bibinfo
		{pages} {115115} (\bibinfo {year} {2020}{\natexlab{b}})}\BibitemShut
	{NoStop}%
	\bibitem [{\citenamefont {Wang}\ \emph {et~al.}(2021)\citenamefont {Wang},
		\citenamefont {Ge}, \citenamefont {Sun}, \citenamefont {Xue}, \citenamefont
		{Jia}, \citenamefont {Guan}, \citenamefont {Yuan}, \citenamefont {Zhang},\
		and\ \citenamefont {Chong}}]{wang.nc.12.3654.2021}%
	\BibitemOpen
	\bibfield  {author} {\bibinfo {author} {\bibfnamefont {Q.}~\bibnamefont
			{Wang}}, \bibinfo {author} {\bibfnamefont {Y.}~\bibnamefont {Ge}}, \bibinfo
		{author} {\bibfnamefont {H.-x.}\ \bibnamefont {Sun}}, \bibinfo {author}
		{\bibfnamefont {H.}~\bibnamefont {Xue}}, \bibinfo {author} {\bibfnamefont
			{D.}~\bibnamefont {Jia}}, \bibinfo {author} {\bibfnamefont {Y.-j.}\
			\bibnamefont {Guan}}, \bibinfo {author} {\bibfnamefont {S.-q.}\ \bibnamefont
			{Yuan}}, \bibinfo {author} {\bibfnamefont {B.}~\bibnamefont {Zhang}},\ and\
		\bibinfo {author} {\bibfnamefont {Y.~D.}\ \bibnamefont {Chong}},\ }\bibfield
	{title} {\bibinfo {title} {Vortex states in an acoustic {Weyl} crystal with a
			topological lattice defect},\ }\href
	{https://doi.org/10.1038/s41467-021-23963-7} {\bibfield  {journal} {\bibinfo
			{journal} {Nat. Commun.}\ }\textbf {\bibinfo {volume} {12}},\ \bibinfo
		{pages} {3654} (\bibinfo {year} {2021})}\BibitemShut {NoStop}%
	\bibitem [{\citenamefont {Ren}\ \emph {et~al.}(2023{\natexlab{a}})\citenamefont
		{Ren}, \citenamefont {Wang}, \citenamefont {Kartashov}, \citenamefont {Li},\
		and\ \citenamefont {Zhang}}]{ren.apl.8.016101.2023}%
	\BibitemOpen
	\bibfield  {author} {\bibinfo {author} {\bibfnamefont {B.}~\bibnamefont
			{Ren}}, \bibinfo {author} {\bibfnamefont {H.}~\bibnamefont {Wang}}, \bibinfo
		{author} {\bibfnamefont {Y.~V.}\ \bibnamefont {Kartashov}}, \bibinfo {author}
		{\bibfnamefont {Y.}~\bibnamefont {Li}},\ and\ \bibinfo {author}
		{\bibfnamefont {Y.}~\bibnamefont {Zhang}},\ }\bibfield  {title} {\bibinfo
		{title} {Nonlinear photonic disclination states},\ }\href
	{https://doi.org/10.1063/5.0126104} {\bibfield  {journal} {\bibinfo
			{journal} {APL Photon.}\ }\textbf {\bibinfo {volume} {8}},\ \bibinfo {pages}
		{016101} (\bibinfo {year} {2023}{\natexlab{a}})}\BibitemShut {NoStop}%
	\bibitem [{\citenamefont {Ren}\ \emph {et~al.}(2023{\natexlab{b}})\citenamefont
		{Ren}, \citenamefont {Arkhipova}, \citenamefont {Zhang}, \citenamefont
		{Kartashov}, \citenamefont {Wang}, \citenamefont {Zhuravitskii},
		\citenamefont {Skryabin}, \citenamefont {Dyakonov}, \citenamefont {Kalinkin},
		\citenamefont {Kulik}, \citenamefont {Kompanets}, \citenamefont {Chekalin},\
		and\ \citenamefont {Zadkov}}]{ren.light.12.194.2023}%
	\BibitemOpen
	\bibfield  {author} {\bibinfo {author} {\bibfnamefont {B.}~\bibnamefont
			{Ren}}, \bibinfo {author} {\bibfnamefont {A.~A.}\ \bibnamefont {Arkhipova}},
		\bibinfo {author} {\bibfnamefont {Y.}~\bibnamefont {Zhang}}, \bibinfo
		{author} {\bibfnamefont {Y.~V.}\ \bibnamefont {Kartashov}}, \bibinfo {author}
		{\bibfnamefont {H.}~\bibnamefont {Wang}}, \bibinfo {author} {\bibfnamefont
			{S.~A.}\ \bibnamefont {Zhuravitskii}}, \bibinfo {author} {\bibfnamefont
			{N.~N.}\ \bibnamefont {Skryabin}}, \bibinfo {author} {\bibfnamefont {I.~V.}\
			\bibnamefont {Dyakonov}}, \bibinfo {author} {\bibfnamefont {A.~A.}\
			\bibnamefont {Kalinkin}}, \bibinfo {author} {\bibfnamefont {S.~P.}\
			\bibnamefont {Kulik}}, \bibinfo {author} {\bibfnamefont {V.~O.}\ \bibnamefont
			{Kompanets}}, \bibinfo {author} {\bibfnamefont {S.~V.}\ \bibnamefont
			{Chekalin}},\ and\ \bibinfo {author} {\bibfnamefont {V.~N.}\ \bibnamefont
			{Zadkov}},\ }\bibfield  {title} {\bibinfo {title} {Observation of nonlinear
			disclination states},\ }\href {https://doi.org/10.1038/s41377-023-01235-x}
	{\bibfield  {journal} {\bibinfo  {journal} {Light Sci. Appl.}\ }\textbf
		{\bibinfo {volume} {12}},\ \bibinfo {pages} {194} (\bibinfo {year}
		{2023}{\natexlab{b}})}\BibitemShut {NoStop}%
	\bibitem [{\citenamefont {Hwang}\ \emph {et~al.}(2024)\citenamefont {Hwang},
		\citenamefont {Kim}, \citenamefont {Kim}, \citenamefont {Yang}, \citenamefont
		{Kivshar},\ and\ \citenamefont {Park}}]{hwang.np.18.286.2024}%
	\BibitemOpen
	\bibfield  {author} {\bibinfo {author} {\bibfnamefont {M.-S.}\ \bibnamefont
			{Hwang}}, \bibinfo {author} {\bibfnamefont {H.-R.}\ \bibnamefont {Kim}},
		\bibinfo {author} {\bibfnamefont {J.}~\bibnamefont {Kim}}, \bibinfo {author}
		{\bibfnamefont {B.-J.}\ \bibnamefont {Yang}}, \bibinfo {author}
		{\bibfnamefont {Y.}~\bibnamefont {Kivshar}},\ and\ \bibinfo {author}
		{\bibfnamefont {H.-G.}\ \bibnamefont {Park}},\ }\bibfield  {title} {\bibinfo
		{title} {Vortex nanolaser based on a photonic disclination cavity},\ }\href
	{https://doi.org/10.1038/s41566-023-01338-2} {\bibfield  {journal} {\bibinfo
			{journal} {Nat. Photon.}\ }\textbf {\bibinfo {volume} {18}},\ \bibinfo
		{pages} {286} (\bibinfo {year} {2024})}\BibitemShut {NoStop}%
	\bibitem [{\citenamefont {Huang}\ \emph {et~al.}(2024)\citenamefont {Huang},
		\citenamefont {Shang}, \citenamefont {Kartashov},\ and\ \citenamefont
		{Ye}}]{huang.nano.13.3495.2024}%
	\BibitemOpen
	\bibfield  {author} {\bibinfo {author} {\bibfnamefont {C.}~\bibnamefont
			{Huang}}, \bibinfo {author} {\bibfnamefont {C.}~\bibnamefont {Shang}},
		\bibinfo {author} {\bibfnamefont {Y.~V.}\ \bibnamefont {Kartashov}},\ and\
		\bibinfo {author} {\bibfnamefont {F.}~\bibnamefont {Ye}},\ }\bibfield
	{title} {\bibinfo {title} {Vortex solitons in topological disclination
			lattices},\ }\href {https://doi.org/doi:10.1515/nanoph-2023-0790} {\bibfield
		{journal} {\bibinfo  {journal} {Nanophoton.}\ }\textbf {\bibinfo {volume}
			{13}},\ \bibinfo {pages} {3495} (\bibinfo {year} {2024})}\BibitemShut
	{NoStop}%
	\bibitem [{\citenamefont {Manna}\ \emph {et~al.}(2022)\citenamefont {Manna},
		\citenamefont {Nandy},\ and\ \citenamefont
		{Roy}}]{manna.prb.105.L201301.2022}%
	\BibitemOpen
	\bibfield  {author} {\bibinfo {author} {\bibfnamefont {S.}~\bibnamefont
			{Manna}}, \bibinfo {author} {\bibfnamefont {S.}~\bibnamefont {Nandy}},\ and\
		\bibinfo {author} {\bibfnamefont {B.}~\bibnamefont {Roy}},\ }\bibfield
	{title} {\bibinfo {title} {Higher-order topological phases on fractal
			lattices},\ }\href {https://doi.org/10.1103/PhysRevB.105.L201301} {\bibfield
		{journal} {\bibinfo  {journal} {Phys. Rev. B}\ }\textbf {\bibinfo {volume}
			{105}},\ \bibinfo {pages} {L201301} (\bibinfo {year} {2022})}\BibitemShut
	{NoStop}%
	\bibitem [{\citenamefont {Li}\ \emph {et~al.}(2022)\citenamefont {Li},
		\citenamefont {Mo}, \citenamefont {Jiang},\ and\ \citenamefont
		{Yang}}]{li.sb.67.2040.2022}%
	\BibitemOpen
	\bibfield  {author} {\bibinfo {author} {\bibfnamefont {J.}~\bibnamefont
			{Li}}, \bibinfo {author} {\bibfnamefont {Q.}~\bibnamefont {Mo}}, \bibinfo
		{author} {\bibfnamefont {J.-H.}\ \bibnamefont {Jiang}},\ and\ \bibinfo
		{author} {\bibfnamefont {Z.}~\bibnamefont {Yang}},\ }\bibfield  {title}
	{\bibinfo {title} {Higher-order topological phase in an acoustic fractal
			lattice},\ }\href {https://doi.org/10.1016/j.scib.2022.09.024} {\bibfield
		{journal} {\bibinfo  {journal} {Sci. Bull.}\ }\textbf {\bibinfo {volume}
			{67}},\ \bibinfo {pages} {2040} (\bibinfo {year} {2022})}\BibitemShut
	{NoStop}%
	\bibitem [{\citenamefont {Zheng}\ \emph {et~al.}(2022)\citenamefont {Zheng},
		\citenamefont {Man}, \citenamefont {Kong}, \citenamefont {Lin}, \citenamefont
		{Duan}, \citenamefont {Chen}, \citenamefont {Yu}, \citenamefont {Jiang},\
		and\ \citenamefont {Xia}}]{zheng.sb.67.2069.2022}%
	\BibitemOpen
	\bibfield  {author} {\bibinfo {author} {\bibfnamefont {S.}~\bibnamefont
			{Zheng}}, \bibinfo {author} {\bibfnamefont {X.}~\bibnamefont {Man}}, \bibinfo
		{author} {\bibfnamefont {Z.-L.}\ \bibnamefont {Kong}}, \bibinfo {author}
		{\bibfnamefont {Z.-K.}\ \bibnamefont {Lin}}, \bibinfo {author} {\bibfnamefont
			{G.}~\bibnamefont {Duan}}, \bibinfo {author} {\bibfnamefont {N.}~\bibnamefont
			{Chen}}, \bibinfo {author} {\bibfnamefont {D.}~\bibnamefont {Yu}}, \bibinfo
		{author} {\bibfnamefont {J.-H.}\ \bibnamefont {Jiang}},\ and\ \bibinfo
		{author} {\bibfnamefont {B.}~\bibnamefont {Xia}},\ }\bibfield  {title}
	{\bibinfo {title} {Observation of fractal higher-order topological states in
			acoustic metamaterials},\ }\href {https://doi.org/10.1016/j.scib.2022.09.020}
	{\bibfield  {journal} {\bibinfo  {journal} {Sci. Bull.}\ }\textbf {\bibinfo
			{volume} {67}},\ \bibinfo {pages} {2069} (\bibinfo {year}
		{2022})}\BibitemShut {NoStop}%
	\bibitem [{\citenamefont {Ren}\ \emph {et~al.}(2023{\natexlab{c}})\citenamefont
		{Ren}, \citenamefont {Kartashov}, \citenamefont {Maczewsky}, \citenamefont
		{Kirsch}, \citenamefont {Wang}, \citenamefont {Szameit}, \citenamefont
		{Heinrich},\ and\ \citenamefont {Zhang}}]{ren.nano.12.3829.2023}%
	\BibitemOpen
	\bibfield  {author} {\bibinfo {author} {\bibfnamefont {B.}~\bibnamefont
			{Ren}}, \bibinfo {author} {\bibfnamefont {Y.~V.}\ \bibnamefont {Kartashov}},
		\bibinfo {author} {\bibfnamefont {L.~J.}\ \bibnamefont {Maczewsky}}, \bibinfo
		{author} {\bibfnamefont {M.~S.}\ \bibnamefont {Kirsch}}, \bibinfo {author}
		{\bibfnamefont {H.}~\bibnamefont {Wang}}, \bibinfo {author} {\bibfnamefont
			{A.}~\bibnamefont {Szameit}}, \bibinfo {author} {\bibfnamefont
			{M.}~\bibnamefont {Heinrich}},\ and\ \bibinfo {author} {\bibfnamefont
			{Y.}~\bibnamefont {Zhang}},\ }\bibfield  {title} {\bibinfo {title} {Theory of
			nonlinear corner states in photonic fractal lattices},\ }\href
	{https://doi.org/10.1515/nanoph-2023-0443} {\bibfield  {journal} {\bibinfo
			{journal} {Nanophoton.}\ }\textbf {\bibinfo {volume} {12}},\ \bibinfo {pages}
		{3829} (\bibinfo {year} {2023}{\natexlab{c}})}\BibitemShut {NoStop}%
	\bibitem [{\citenamefont {Li}\ \emph {et~al.}(2023)\citenamefont {Li},
		\citenamefont {Li}, \citenamefont {Yan}, \citenamefont {Li}, \citenamefont
		{Gong},\ and\ \citenamefont {Li}}]{li.light.12.262.2023}%
	\BibitemOpen
	\bibfield  {author} {\bibinfo {author} {\bibfnamefont {M.}~\bibnamefont
			{Li}}, \bibinfo {author} {\bibfnamefont {C.}~\bibnamefont {Li}}, \bibinfo
		{author} {\bibfnamefont {L.}~\bibnamefont {Yan}}, \bibinfo {author}
		{\bibfnamefont {Q.}~\bibnamefont {Li}}, \bibinfo {author} {\bibfnamefont
			{Q.}~\bibnamefont {Gong}},\ and\ \bibinfo {author} {\bibfnamefont
			{Y.}~\bibnamefont {Li}},\ }\bibfield  {title} {\bibinfo {title} {Fractal
			photonic anomalous {Floquet} topological insulators to generate multiple
			quantum chiral edge states},\ }\href
	{https://doi.org/10.1038/s41377-023-01307-y} {\bibfield  {journal} {\bibinfo
			{journal} {Light Sci. Appl.}\ }\textbf {\bibinfo {volume} {12}},\ \bibinfo
		{pages} {262} (\bibinfo {year} {2023})}\BibitemShut {NoStop}%
	\bibitem [{\citenamefont {Eek}\ \emph {et~al.}(2025)\citenamefont {Eek},
		\citenamefont {Osseweijer},\ and\ \citenamefont
		{Morais~Smith}}]{eek.prl.134.246601.2025}%
	\BibitemOpen
	\bibfield  {author} {\bibinfo {author} {\bibfnamefont {L.}~\bibnamefont
			{Eek}}, \bibinfo {author} {\bibfnamefont {Z.~F.}\ \bibnamefont
			{Osseweijer}},\ and\ \bibinfo {author} {\bibfnamefont {C.}~\bibnamefont
			{Morais~Smith}},\ }\bibfield  {title} {\bibinfo {title} {Fractality-induced
			topology},\ }\href {https://doi.org/10.1103/jcrl-9dz6} {\bibfield  {journal}
		{\bibinfo  {journal} {Phys. Rev. Lett.}\ }\textbf {\bibinfo {volume} {134}},\
		\bibinfo {pages} {246601} (\bibinfo {year} {2025})}\BibitemShut {NoStop}%
	\bibitem [{\citenamefont {Kartashov}\ \emph {et~al.}(2022)\citenamefont
		{Kartashov}, \citenamefont {Arkhipova}, \citenamefont {Zhuravitskii},
		\citenamefont {Skryabin}, \citenamefont {Dyakonov}, \citenamefont {Kalinkin},
		\citenamefont {Kulik}, \citenamefont {Kompanets}, \citenamefont {Chekalin},
		\citenamefont {Torner},\ and\ \citenamefont
		{Zadkov}}]{kartashov.prl.128.093901.2022}%
	\BibitemOpen
	\bibfield  {author} {\bibinfo {author} {\bibfnamefont {Y.~V.}\ \bibnamefont
			{Kartashov}}, \bibinfo {author} {\bibfnamefont {A.~A.}\ \bibnamefont
			{Arkhipova}}, \bibinfo {author} {\bibfnamefont {S.~A.}\ \bibnamefont
			{Zhuravitskii}}, \bibinfo {author} {\bibfnamefont {N.~N.}\ \bibnamefont
			{Skryabin}}, \bibinfo {author} {\bibfnamefont {I.~V.}\ \bibnamefont
			{Dyakonov}}, \bibinfo {author} {\bibfnamefont {A.~A.}\ \bibnamefont
			{Kalinkin}}, \bibinfo {author} {\bibfnamefont {S.~P.}\ \bibnamefont {Kulik}},
		\bibinfo {author} {\bibfnamefont {V.~O.}\ \bibnamefont {Kompanets}}, \bibinfo
		{author} {\bibfnamefont {S.~V.}\ \bibnamefont {Chekalin}}, \bibinfo {author}
		{\bibfnamefont {L.}~\bibnamefont {Torner}},\ and\ \bibinfo {author}
		{\bibfnamefont {V.~N.}\ \bibnamefont {Zadkov}},\ }\bibfield  {title}
	{\bibinfo {title} {Observation of edge solitons in topological trimer
			arrays},\ }\href {https://doi.org/10.1103/PhysRevLett.128.093901} {\bibfield
		{journal} {\bibinfo  {journal} {Phys. Rev. Lett.}\ }\textbf {\bibinfo
			{volume} {128}},\ \bibinfo {pages} {093901} (\bibinfo {year}
		{2022})}\BibitemShut {NoStop}%
	\bibitem [{\citenamefont {Chen}\ \emph {et~al.}(2025)\citenamefont {Chen},
		\citenamefont {Yan}, \citenamefont {Liu}, \citenamefont {Cheng},
		\citenamefont {Lu}, \citenamefont {Tan},\ and\ \citenamefont
		{Chen}}]{chen.sb.70.1605.2025}%
	\BibitemOpen
	\bibfield  {author} {\bibinfo {author} {\bibfnamefont {R.}~\bibnamefont
			{Chen}}, \bibinfo {author} {\bibfnamefont {W.}~\bibnamefont {Yan}}, \bibinfo
		{author} {\bibfnamefont {W.}~\bibnamefont {Liu}}, \bibinfo {author}
		{\bibfnamefont {W.}~\bibnamefont {Cheng}}, \bibinfo {author} {\bibfnamefont
			{Q.}~\bibnamefont {Lu}}, \bibinfo {author} {\bibfnamefont {Y.}~\bibnamefont
			{Tan}},\ and\ \bibinfo {author} {\bibfnamefont {F.}~\bibnamefont {Chen}},\
	}\bibfield  {title} {\bibinfo {title} {Nonlinear tuning of multiple
			topological edge states in photovoltaic photonic lattices},\ }\href
	{https://doi.org/https://doi.org/10.1016/j.scib.2025.03.018} {\bibfield
		{journal} {\bibinfo  {journal} {Sci. Bull.}\ }\textbf {\bibinfo {volume}
			{70}},\ \bibinfo {pages} {1605} (\bibinfo {year} {2025})}\BibitemShut
	{NoStop}%
	\bibitem [{\citenamefont {Du}\ \emph {et~al.}(2025)\citenamefont {Du},
		\citenamefont {Zhao}, \citenamefont {Li}, \citenamefont {Wang}, \citenamefont
		{Li}, \citenamefont {Wu}, \citenamefont {Liu}, \citenamefont {Zhang},
		\citenamefont {Xiao}, \citenamefont {Jia},\ and\ \citenamefont
		{Ma}}]{du.light.14.296.2025}%
	\BibitemOpen
	\bibfield  {author} {\bibinfo {author} {\bibfnamefont {H.}~\bibnamefont
			{Du}}, \bibinfo {author} {\bibfnamefont {H.}~\bibnamefont {Zhao}}, \bibinfo
		{author} {\bibfnamefont {Y.}~\bibnamefont {Li}}, \bibinfo {author}
		{\bibfnamefont {Y.}~\bibnamefont {Wang}}, \bibinfo {author} {\bibfnamefont
			{R.}~\bibnamefont {Li}}, \bibinfo {author} {\bibfnamefont {J.}~\bibnamefont
			{Wu}}, \bibinfo {author} {\bibfnamefont {W.}~\bibnamefont {Liu}}, \bibinfo
		{author} {\bibfnamefont {Y.}~\bibnamefont {Zhang}}, \bibinfo {author}
		{\bibfnamefont {L.}~\bibnamefont {Xiao}}, \bibinfo {author} {\bibfnamefont
			{S.}~\bibnamefont {Jia}},\ and\ \bibinfo {author} {\bibfnamefont
			{J.}~\bibnamefont {Ma}},\ }\bibfield  {title} {\bibinfo {title} {Observation
			of nonlinear edge states in an interacting atomic trimer array},\ }\href
	{https://doi.org/10.1038/s41377-025-01997-6} {\bibfield  {journal} {\bibinfo
			{journal} {Light Sci. Appl.}\ }\textbf {\bibinfo {volume} {14}},\ \bibinfo
		{pages} {296} (\bibinfo {year} {2025})}\BibitemShut {NoStop}%
	\bibitem [{\citenamefont {Smirnova}\ \emph {et~al.}(2020)\citenamefont
		{Smirnova}, \citenamefont {Leykam}, \citenamefont {Chong},\ and\
		\citenamefont {Kivshar}}]{smirnova.apr.7.021306.2020}%
	\BibitemOpen
	\bibfield  {author} {\bibinfo {author} {\bibfnamefont {D.}~\bibnamefont
			{Smirnova}}, \bibinfo {author} {\bibfnamefont {D.}~\bibnamefont {Leykam}},
		\bibinfo {author} {\bibfnamefont {Y.}~\bibnamefont {Chong}},\ and\ \bibinfo
		{author} {\bibfnamefont {Y.}~\bibnamefont {Kivshar}},\ }\bibfield  {title}
	{\bibinfo {title} {Nonlinear topological photonics},\ }\href
	{https://doi.org/10.1063/1.5142397} {\bibfield  {journal} {\bibinfo
			{journal} {Appl. Phys. Rev.}\ }\textbf {\bibinfo {volume} {7}},\ \bibinfo
		{pages} {021306} (\bibinfo {year} {2020})}\BibitemShut {NoStop}%
	\bibitem [{\citenamefont {Szameit}\ and\ \citenamefont
		{Rechtsman}(2024)}]{szameit.np.20.905.2024}%
	\BibitemOpen
	\bibfield  {author} {\bibinfo {author} {\bibfnamefont {A.}~\bibnamefont
			{Szameit}}\ and\ \bibinfo {author} {\bibfnamefont {M.~C.}\ \bibnamefont
			{Rechtsman}},\ }\bibfield  {title} {\bibinfo {title} {Discrete nonlinear
			topological photonics},\ }\href {https://doi.org/10.1038/s41567-024-02454-8}
	{\bibfield  {journal} {\bibinfo  {journal} {Nat. Phys.}\ }\textbf {\bibinfo
			{volume} {20}},\ \bibinfo {pages} {905} (\bibinfo {year} {2024})}\BibitemShut
	{NoStop}%
	\bibitem [{\citenamefont {Maczewsky}\ \emph {et~al.}(2020)\citenamefont
		{Maczewsky}, \citenamefont {Heinrich}, \citenamefont {Kremer}, \citenamefont
		{Ivanov}, \citenamefont {Ehrhardt}, \citenamefont {Martinez}, \citenamefont
		{Kartashov}, \citenamefont {Konotop}, \citenamefont {Torner}, \citenamefont
		{Bauer},\ and\ \citenamefont {Szameit}}]{maczewsky.science.370.701.2020}%
	\BibitemOpen
	\bibfield  {author} {\bibinfo {author} {\bibfnamefont {L.~J.}\ \bibnamefont
			{Maczewsky}}, \bibinfo {author} {\bibfnamefont {M.}~\bibnamefont {Heinrich}},
		\bibinfo {author} {\bibfnamefont {M.}~\bibnamefont {Kremer}}, \bibinfo
		{author} {\bibfnamefont {S.~K.}\ \bibnamefont {Ivanov}}, \bibinfo {author}
		{\bibfnamefont {M.}~\bibnamefont {Ehrhardt}}, \bibinfo {author}
		{\bibfnamefont {F.}~\bibnamefont {Martinez}}, \bibinfo {author}
		{\bibfnamefont {Y.~V.}\ \bibnamefont {Kartashov}}, \bibinfo {author}
		{\bibfnamefont {V.~V.}\ \bibnamefont {Konotop}}, \bibinfo {author}
		{\bibfnamefont {L.}~\bibnamefont {Torner}}, \bibinfo {author} {\bibfnamefont
			{D.}~\bibnamefont {Bauer}},\ and\ \bibinfo {author} {\bibfnamefont
			{A.}~\bibnamefont {Szameit}},\ }\bibfield  {title} {\bibinfo {title}
		{Nonlinearity-induced photonic topological insulator},\ }\href
	{https://doi.org/10.1126/science.abd2033} {\bibfield  {journal} {\bibinfo
			{journal} {Science}\ }\textbf {\bibinfo {volume} {370}},\ \bibinfo {pages}
		{701} (\bibinfo {year} {2020})}\BibitemShut {NoStop}%
	\bibitem [{\citenamefont {Mukherjee}\ and\ \citenamefont
		{Rechtsman}(2020)}]{mukherjee.science.368.856.2020}%
	\BibitemOpen
	\bibfield  {author} {\bibinfo {author} {\bibfnamefont {S.}~\bibnamefont
			{Mukherjee}}\ and\ \bibinfo {author} {\bibfnamefont {M.~C.}\ \bibnamefont
			{Rechtsman}},\ }\bibfield  {title} {\bibinfo {title} {Observation of
			{F}loquet solitons in a topological bandgap},\ }\href
	{https://doi.org/10.1126/science.aba8725} {\bibfield  {journal} {\bibinfo
			{journal} {Science}\ }\textbf {\bibinfo {volume} {368}},\ \bibinfo {pages}
		{856} (\bibinfo {year} {2020})}\BibitemShut {NoStop}%
	\bibitem [{\citenamefont {Sone}\ \emph {et~al.}(2024)\citenamefont {Sone},
		\citenamefont {Ezawa}, \citenamefont {Ashida}, \citenamefont {Yoshioka},\
		and\ \citenamefont {Sagawa}}]{sone.np.20.1164.2024}%
	\BibitemOpen
	\bibfield  {author} {\bibinfo {author} {\bibfnamefont {K.}~\bibnamefont
			{Sone}}, \bibinfo {author} {\bibfnamefont {M.}~\bibnamefont {Ezawa}},
		\bibinfo {author} {\bibfnamefont {Y.}~\bibnamefont {Ashida}}, \bibinfo
		{author} {\bibfnamefont {N.}~\bibnamefont {Yoshioka}},\ and\ \bibinfo
		{author} {\bibfnamefont {T.}~\bibnamefont {Sagawa}},\ }\bibfield  {title}
	{\bibinfo {title} {Nonlinearity-induced topological phase transition
			characterized by the nonlinear {Chern} number},\ }\href
	{https://doi.org/10.1038/s41567-024-02451-x} {\bibfield  {journal} {\bibinfo
			{journal} {Nat. Phys.}\ }\textbf {\bibinfo {volume} {20}},\ \bibinfo {pages}
		{1164} (\bibinfo {year} {2024})}\BibitemShut {NoStop}%
	\bibitem [{\citenamefont {Kirsch}\ \emph {et~al.}(2021)\citenamefont {Kirsch},
		\citenamefont {Zhang}, \citenamefont {Kremer}, \citenamefont {Maczewsky},
		\citenamefont {Ivanov}, \citenamefont {Kartashov}, \citenamefont {Torner},
		\citenamefont {Bauer}, \citenamefont {Szameit},\ and\ \citenamefont
		{Heinrich}}]{kirsch.np.17.995.2021}%
	\BibitemOpen
	\bibfield  {author} {\bibinfo {author} {\bibfnamefont {M.~S.}\ \bibnamefont
			{Kirsch}}, \bibinfo {author} {\bibfnamefont {Y.}~\bibnamefont {Zhang}},
		\bibinfo {author} {\bibfnamefont {M.}~\bibnamefont {Kremer}}, \bibinfo
		{author} {\bibfnamefont {L.~J.}\ \bibnamefont {Maczewsky}}, \bibinfo {author}
		{\bibfnamefont {S.~K.}\ \bibnamefont {Ivanov}}, \bibinfo {author}
		{\bibfnamefont {Y.~V.}\ \bibnamefont {Kartashov}}, \bibinfo {author}
		{\bibfnamefont {L.}~\bibnamefont {Torner}}, \bibinfo {author} {\bibfnamefont
			{D.}~\bibnamefont {Bauer}}, \bibinfo {author} {\bibfnamefont
			{A.}~\bibnamefont {Szameit}},\ and\ \bibinfo {author} {\bibfnamefont
			{M.}~\bibnamefont {Heinrich}},\ }\bibfield  {title} {\bibinfo {title}
		{Nonlinear second-order photonic topological insulators},\ }\href
	{https://doi.org/10.1038/s41567-021-01275-3} {\bibfield  {journal} {\bibinfo
			{journal} {Nat. Phys.}\ }\textbf {\bibinfo {volume} {17}},\ \bibinfo {pages}
		{995} (\bibinfo {year} {2021})}\BibitemShut {NoStop}%
	\bibitem [{\citenamefont {Hu}\ \emph {et~al.}(2021)\citenamefont {Hu},
		\citenamefont {Bongiovanni}, \citenamefont {Juki\'c}, \citenamefont
		{Jajti\'c}, \citenamefont {Xia}, \citenamefont {Song}, \citenamefont {Xu},
		\citenamefont {Morandotti}, \citenamefont {Buljan},\ and\ \citenamefont
		{Chen}}]{hu.light.10.164.2021}%
	\BibitemOpen
	\bibfield  {author} {\bibinfo {author} {\bibfnamefont {Z.}~\bibnamefont
			{Hu}}, \bibinfo {author} {\bibfnamefont {D.}~\bibnamefont {Bongiovanni}},
		\bibinfo {author} {\bibfnamefont {D.}~\bibnamefont {Juki\'c}}, \bibinfo
		{author} {\bibfnamefont {E.}~\bibnamefont {Jajti\'c}}, \bibinfo {author}
		{\bibfnamefont {S.}~\bibnamefont {Xia}}, \bibinfo {author} {\bibfnamefont
			{D.}~\bibnamefont {Song}}, \bibinfo {author} {\bibfnamefont {J.}~\bibnamefont
			{Xu}}, \bibinfo {author} {\bibfnamefont {R.}~\bibnamefont {Morandotti}},
		\bibinfo {author} {\bibfnamefont {H.}~\bibnamefont {Buljan}},\ and\ \bibinfo
		{author} {\bibfnamefont {Z.}~\bibnamefont {Chen}},\ }\bibfield  {title}
	{\bibinfo {title} {Nonlinear control of photonic higher-order topological
			bound states in the continuum},\ }\href
	{https://doi.org/10.1038/s41377-021-00607-5} {\bibfield  {journal} {\bibinfo
			{journal} {Light Sci. Appl.}\ }\textbf {\bibinfo {volume} {10}},\ \bibinfo
		{pages} {164} (\bibinfo {year} {2021})}\BibitemShut {NoStop}%
	\bibitem [{\citenamefont {Ezawa}(2021)}]{ezawa.prb.104.235420.2021}%
	\BibitemOpen
	\bibfield  {author} {\bibinfo {author} {\bibfnamefont {M.}~\bibnamefont
			{Ezawa}},\ }\bibfield  {title} {\bibinfo {title} {Nonlinearity-induced
			transition in the nonlinear {Su-Schrieffer-Heeger} model and a nonlinear
			higher-order topological system},\ }\href
	{https://doi.org/10.1103/PhysRevB.104.235420} {\bibfield  {journal} {\bibinfo
			{journal} {Phys. Rev. B}\ }\textbf {\bibinfo {volume} {104}},\ \bibinfo
		{pages} {235420} (\bibinfo {year} {2021})}\BibitemShut {NoStop}%
	\bibitem [{\citenamefont {Zhong}\ \emph {et~al.}(2024)\citenamefont {Zhong},
		\citenamefont {Kompanets}, \citenamefont {Zhang}, \citenamefont {Kartashov},
		\citenamefont {Cao}, \citenamefont {Li}, \citenamefont {Zhuravitskii},
		\citenamefont {Skryabin}, \citenamefont {Dyakonov}, \citenamefont {Kalinkin},
		\citenamefont {Kulik}, \citenamefont {Chekalin},\ and\ \citenamefont
		{Zadkov}}]{zhong.light.13.264.2024}%
	\BibitemOpen
	\bibfield  {author} {\bibinfo {author} {\bibfnamefont {H.}~\bibnamefont
			{Zhong}}, \bibinfo {author} {\bibfnamefont {V.~O.}\ \bibnamefont
			{Kompanets}}, \bibinfo {author} {\bibfnamefont {Y.}~\bibnamefont {Zhang}},
		\bibinfo {author} {\bibfnamefont {Y.~V.}\ \bibnamefont {Kartashov}}, \bibinfo
		{author} {\bibfnamefont {M.}~\bibnamefont {Cao}}, \bibinfo {author}
		{\bibfnamefont {Y.}~\bibnamefont {Li}}, \bibinfo {author} {\bibfnamefont
			{S.~A.}\ \bibnamefont {Zhuravitskii}}, \bibinfo {author} {\bibfnamefont
			{N.~N.}\ \bibnamefont {Skryabin}}, \bibinfo {author} {\bibfnamefont {I.~V.}\
			\bibnamefont {Dyakonov}}, \bibinfo {author} {\bibfnamefont {A.~A.}\
			\bibnamefont {Kalinkin}}, \bibinfo {author} {\bibfnamefont {S.~P.}\
			\bibnamefont {Kulik}}, \bibinfo {author} {\bibfnamefont {S.~V.}\ \bibnamefont
			{Chekalin}},\ and\ \bibinfo {author} {\bibfnamefont {V.~N.}\ \bibnamefont
			{Zadkov}},\ }\bibfield  {title} {\bibinfo {title} {Observation of nonlinear
			fractal higher order topological insulator},\ }\href
	{https://doi.org/10.1038/s41377-024-01611-1} {\bibfield  {journal} {\bibinfo
			{journal} {Light Sci. Appl.}\ }\textbf {\bibinfo {volume} {13}},\ \bibinfo
		{pages} {264} (\bibinfo {year} {2024})}\BibitemShut {NoStop}%
	\bibitem [{\citenamefont {Kompanets}\ \emph {et~al.}(2025)\citenamefont
		{Kompanets}, \citenamefont {Feng}, \citenamefont {Zhang}, \citenamefont
		{Kartashov}, \citenamefont {Li}, \citenamefont {Zhuravitskii}, \citenamefont
		{Skryabin}, \citenamefont {Kireev}, \citenamefont {Dyakonov}, \citenamefont
		{Kalinkin}, \citenamefont {Shang}, \citenamefont {Kulik}, \citenamefont
		{Chekalin},\ and\ \citenamefont {Zadkov}}]{kompanets.am.37.2500556.2025}%
	\BibitemOpen
	\bibfield  {author} {\bibinfo {author} {\bibfnamefont {V.~O.}\ \bibnamefont
			{Kompanets}}, \bibinfo {author} {\bibfnamefont {S.}~\bibnamefont {Feng}},
		\bibinfo {author} {\bibfnamefont {Y.}~\bibnamefont {Zhang}}, \bibinfo
		{author} {\bibfnamefont {Y.~V.}\ \bibnamefont {Kartashov}}, \bibinfo {author}
		{\bibfnamefont {Y.}~\bibnamefont {Li}}, \bibinfo {author} {\bibfnamefont
			{S.~A.}\ \bibnamefont {Zhuravitskii}}, \bibinfo {author} {\bibfnamefont
			{N.~N.}\ \bibnamefont {Skryabin}}, \bibinfo {author} {\bibfnamefont {A.~V.}\
			\bibnamefont {Kireev}}, \bibinfo {author} {\bibfnamefont {I.~V.}\
			\bibnamefont {Dyakonov}}, \bibinfo {author} {\bibfnamefont {A.~A.}\
			\bibnamefont {Kalinkin}}, \bibinfo {author} {\bibfnamefont {C.}~\bibnamefont
			{Shang}}, \bibinfo {author} {\bibfnamefont {S.~P.}\ \bibnamefont {Kulik}},
		\bibinfo {author} {\bibfnamefont {S.~V.}\ \bibnamefont {Chekalin}},\ and\
		\bibinfo {author} {\bibfnamefont {V.~N.}\ \bibnamefont {Zadkov}},\ }\bibfield
	{title} {\bibinfo {title} {Observation of nonlinear topological corner
			states originating from different spectral charges},\ }\href
	{https://doi.org/10.1002/adma.202500556} {\bibfield  {journal} {\bibinfo
			{journal} {Adv. Mater.}\ }\textbf {\bibinfo {volume} {37}},\ \bibinfo {pages}
		{2500556} (\bibinfo {year} {2025})}\BibitemShut {NoStop}%
	\bibitem [{\citenamefont {Banerjee}\ \emph {et~al.}(2020)\citenamefont
		{Banerjee}, \citenamefont {Mandal},\ and\ \citenamefont
		{Liew}}]{banerjee.prl.124.063901.2020}%
	\BibitemOpen
	\bibfield  {author} {\bibinfo {author} {\bibfnamefont {R.}~\bibnamefont
			{Banerjee}}, \bibinfo {author} {\bibfnamefont {S.}~\bibnamefont {Mandal}},\
		and\ \bibinfo {author} {\bibfnamefont {T.~C.~H.}\ \bibnamefont {Liew}},\
	}\bibfield  {title} {\bibinfo {title} {Coupling between exciton-polariton
			corner modes through edge states},\ }\href
	{https://doi.org/10.1103/PhysRevLett.124.063901} {\bibfield  {journal}
		{\bibinfo  {journal} {Phys. Rev. Lett.}\ }\textbf {\bibinfo {volume} {124}},\
		\bibinfo {pages} {063901} (\bibinfo {year} {2020})}\BibitemShut {NoStop}%
	\bibitem [{\citenamefont {Zhang}\ \emph {et~al.}(2020)\citenamefont {Zhang},
		\citenamefont {Kartashov}, \citenamefont {Torner}, \citenamefont {Li},\ and\
		\citenamefont {Ferrando}}]{zhang.ol.45.4710.2020}%
	\BibitemOpen
	\bibfield  {author} {\bibinfo {author} {\bibfnamefont {Y.}~\bibnamefont
			{Zhang}}, \bibinfo {author} {\bibfnamefont {Y.~V.}\ \bibnamefont
			{Kartashov}}, \bibinfo {author} {\bibfnamefont {L.}~\bibnamefont {Torner}},
		\bibinfo {author} {\bibfnamefont {Y.}~\bibnamefont {Li}},\ and\ \bibinfo
		{author} {\bibfnamefont {A.}~\bibnamefont {Ferrando}},\ }\bibfield  {title}
	{\bibinfo {title} {Nonlinear higher-order polariton topological insulator},\
	}\href {https://doi.org/10.1364/OL.396039} {\bibfield  {journal} {\bibinfo
			{journal} {Opt. Lett.}\ }\textbf {\bibinfo {volume} {45}},\ \bibinfo {pages}
		{4710} (\bibinfo {year} {2020})}\BibitemShut {NoStop}%
	\bibitem [{\citenamefont {Wu}\ \emph {et~al.}(2023)\citenamefont {Wu},
		\citenamefont {Ghosh}, \citenamefont {Gan}, \citenamefont {Shi},
		\citenamefont {Mandal}, \citenamefont {Sun}, \citenamefont {Zhang},
		\citenamefont {Liew}, \citenamefont {Su},\ and\ \citenamefont
		{Xiong}}]{wu.sa.9.4322.2023}%
	\BibitemOpen
	\bibfield  {author} {\bibinfo {author} {\bibfnamefont {J.}~\bibnamefont
			{Wu}}, \bibinfo {author} {\bibfnamefont {S.}~\bibnamefont {Ghosh}}, \bibinfo
		{author} {\bibfnamefont {Y.}~\bibnamefont {Gan}}, \bibinfo {author}
		{\bibfnamefont {Y.}~\bibnamefont {Shi}}, \bibinfo {author} {\bibfnamefont
			{S.}~\bibnamefont {Mandal}}, \bibinfo {author} {\bibfnamefont
			{H.}~\bibnamefont {Sun}}, \bibinfo {author} {\bibfnamefont {B.}~\bibnamefont
			{Zhang}}, \bibinfo {author} {\bibfnamefont {T.~C.~H.}\ \bibnamefont {Liew}},
		\bibinfo {author} {\bibfnamefont {R.}~\bibnamefont {Su}},\ and\ \bibinfo
		{author} {\bibfnamefont {Q.}~\bibnamefont {Xiong}},\ }\bibfield  {title}
	{\bibinfo {title} {Higher-order topological polariton corner state lasing},\
	}\href {https://doi.org/10.1126/sciadv.adg4322} {\bibfield  {journal}
		{\bibinfo  {journal} {Sci. Adv.}\ }\textbf {\bibinfo {volume} {9}},\ \bibinfo
		{pages} {eadg4322} (\bibinfo {year} {2023})}\BibitemShut {NoStop}%
	\bibitem [{\citenamefont {Schneider}\ \emph {et~al.}(2024)\citenamefont
		{Schneider}, \citenamefont {Gao}, \citenamefont {Zentgraf}, \citenamefont
		{Schumacher},\ and\ \citenamefont {Ma}}]{schneider.nano.13.509.2024}%
	\BibitemOpen
	\bibfield  {author} {\bibinfo {author} {\bibfnamefont {T.}~\bibnamefont
			{Schneider}}, \bibinfo {author} {\bibfnamefont {W.}~\bibnamefont {Gao}},
		\bibinfo {author} {\bibfnamefont {T.}~\bibnamefont {Zentgraf}}, \bibinfo
		{author} {\bibfnamefont {S.}~\bibnamefont {Schumacher}},\ and\ \bibinfo
		{author} {\bibfnamefont {X.}~\bibnamefont {Ma}},\ }\bibfield  {title}
	{\bibinfo {title} {Topological edge and corner states in coupled wave
			lattices in nonlinear polariton condensates},\ }\href
	{https://doi.org/10.1515/nanoph-2023-0556} {\bibfield  {journal} {\bibinfo
			{journal} {Nanophoton.}\ }\textbf {\bibinfo {volume} {13}},\ \bibinfo {pages}
		{509} (\bibinfo {year} {2024})}\BibitemShut {NoStop}%
	\bibitem [{\citenamefont {Bennenhei}\ \emph {et~al.}(2024)\citenamefont
		{Bennenhei}, \citenamefont {Shan}, \citenamefont {Struve}, \citenamefont
		{Kunte}, \citenamefont {Eilenberger}, \citenamefont {Ohmer}, \citenamefont
		{Fischer}, \citenamefont {Schumacher}, \citenamefont {Ma}, \citenamefont
		{Schneider},\ and\ \citenamefont {Esmann}}]{bennenhei.acs.11.3046.2024}%
	\BibitemOpen
	\bibfield  {author} {\bibinfo {author} {\bibfnamefont {C.}~\bibnamefont
			{Bennenhei}}, \bibinfo {author} {\bibfnamefont {H.}~\bibnamefont {Shan}},
		\bibinfo {author} {\bibfnamefont {M.}~\bibnamefont {Struve}}, \bibinfo
		{author} {\bibfnamefont {N.}~\bibnamefont {Kunte}}, \bibinfo {author}
		{\bibfnamefont {F.}~\bibnamefont {Eilenberger}}, \bibinfo {author}
		{\bibfnamefont {J.}~\bibnamefont {Ohmer}}, \bibinfo {author} {\bibfnamefont
			{U.}~\bibnamefont {Fischer}}, \bibinfo {author} {\bibfnamefont
			{S.}~\bibnamefont {Schumacher}}, \bibinfo {author} {\bibfnamefont
			{X.}~\bibnamefont {Ma}}, \bibinfo {author} {\bibfnamefont {C.}~\bibnamefont
			{Schneider}},\ and\ \bibinfo {author} {\bibfnamefont {M.}~\bibnamefont
			{Esmann}},\ }\bibfield  {title} {\bibinfo {title} {Organic room-temperature
			polariton condensate in a higher-order topological lattice},\ }\href
	{https://doi.org/10.1021/acsphotonics.4c00268} {\bibfield  {journal}
		{\bibinfo  {journal} {ACS Photon.}\ }\textbf {\bibinfo {volume} {11}},\
		\bibinfo {pages} {3046} (\bibinfo {year} {2024})}\BibitemShut {NoStop}%
	\bibitem [{\citenamefont {Szameit}\ and\ \citenamefont
		{Nolte}(2010)}]{szameit.jpb.43.163001.2010}%
	\BibitemOpen
	\bibfield  {author} {\bibinfo {author} {\bibfnamefont {A.}~\bibnamefont
			{Szameit}}\ and\ \bibinfo {author} {\bibfnamefont {S.}~\bibnamefont
			{Nolte}},\ }\bibfield  {title} {\bibinfo {title} {Discrete optics in
			femtosecond-laser-written photonic structures},\ }\href
	{https://doi.org/10.1088/0953-4075/43/16/163001} {\bibfield  {journal}
		{\bibinfo  {journal} {J. Phys. B: At. Mol. Opt. Phys.}\ }\textbf {\bibinfo
			{volume} {43}},\ \bibinfo {pages} {163001} (\bibinfo {year}
		{2010})}\BibitemShut {NoStop}%
	\bibitem [{\citenamefont {Skryabin}\ \emph {et~al.}(2024)\citenamefont
		{Skryabin}, \citenamefont {Zhuravitskii}, \citenamefont {Dyakonov},
		\citenamefont {Straupe}, \citenamefont {Kalinkin},\ and\ \citenamefont
		{Kulik}}]{skryabin.prap.22.064079.2024}%
	\BibitemOpen
	\bibfield  {author} {\bibinfo {author} {\bibfnamefont {N.~N.}\ \bibnamefont
			{Skryabin}}, \bibinfo {author} {\bibfnamefont {S.~A.}\ \bibnamefont
			{Zhuravitskii}}, \bibinfo {author} {\bibfnamefont {I.~V.}\ \bibnamefont
			{Dyakonov}}, \bibinfo {author} {\bibfnamefont {S.~S.}\ \bibnamefont
			{Straupe}}, \bibinfo {author} {\bibfnamefont {A.~A.}\ \bibnamefont
			{Kalinkin}},\ and\ \bibinfo {author} {\bibfnamefont {S.~P.}\ \bibnamefont
			{Kulik}},\ }\bibfield  {title} {\bibinfo {title} {Femtosecond-laser-written
			low-loss multiscan waveguides in fused silica},\ }\href
	{https://doi.org/10.1103/PhysRevApplied.22.064079} {\bibfield  {journal}
		{\bibinfo  {journal} {Phys. Rev. Appl.}\ }\textbf {\bibinfo {volume} {22}},\
		\bibinfo {pages} {064079} (\bibinfo {year} {2024})}\BibitemShut {NoStop}%
	\bibitem [{\citenamefont {Yan}\ \emph {et~al.}(2024)\citenamefont {Yan},
		\citenamefont {Zhang},\ and\ \citenamefont {Chen}}]{yan.npjn.1.40.2024}%
	\BibitemOpen
	\bibfield  {author} {\bibinfo {author} {\bibfnamefont {W.}~\bibnamefont
			{Yan}}, \bibinfo {author} {\bibfnamefont {B.}~\bibnamefont {Zhang}},\ and\
		\bibinfo {author} {\bibfnamefont {F.}~\bibnamefont {Chen}},\ }\bibfield
	{title} {\bibinfo {title} {Photonic topological insulators in femtosecond
			laser direct-written waveguides},\ }\href
	{https://doi.org/10.1038/s44310-024-00040-7} {\bibfield  {journal} {\bibinfo
			{journal} {npj Nanophoton.}\ }\textbf {\bibinfo {volume} {1}},\ \bibinfo
		{pages} {40} (\bibinfo {year} {2024})}\BibitemShut {NoStop}%
	\bibitem [{\citenamefont {Yang}\ \emph {et~al.}(2020)\citenamefont {Yang},
		\citenamefont {Lustig}, \citenamefont {Lumer},\ and\ \citenamefont
		{Segev}}]{yang.light.9.128.2020}%
	\BibitemOpen
	\bibfield  {author} {\bibinfo {author} {\bibfnamefont {Z.}~\bibnamefont
			{Yang}}, \bibinfo {author} {\bibfnamefont {E.}~\bibnamefont {Lustig}},
		\bibinfo {author} {\bibfnamefont {Y.}~\bibnamefont {Lumer}},\ and\ \bibinfo
		{author} {\bibfnamefont {M.}~\bibnamefont {Segev}},\ }\bibfield  {title}
	{\bibinfo {title} {Photonic {F}loquet topological insulators in a fractal
			lattice},\ }\href {https://doi.org/10.1038/s41377-020-00354-z} {\bibfield
		{journal} {\bibinfo  {journal} {Light Sci. Appl.}\ }\textbf {\bibinfo
			{volume} {9}},\ \bibinfo {pages} {128} (\bibinfo {year} {2020})}\BibitemShut
	{NoStop}%
	\bibitem [{\citenamefont {Biesenthal}\ \emph {et~al.}(2022)\citenamefont
		{Biesenthal}, \citenamefont {Maczewsky}, \citenamefont {Yang}, \citenamefont
		{Kremer}, \citenamefont {Segev}, \citenamefont {Szameit},\ and\ \citenamefont
		{Heinrich}}]{biesenthal.science.376.1114.2022}%
	\BibitemOpen
	\bibfield  {author} {\bibinfo {author} {\bibfnamefont {T.}~\bibnamefont
			{Biesenthal}}, \bibinfo {author} {\bibfnamefont {L.~J.}\ \bibnamefont
			{Maczewsky}}, \bibinfo {author} {\bibfnamefont {Z.}~\bibnamefont {Yang}},
		\bibinfo {author} {\bibfnamefont {M.}~\bibnamefont {Kremer}}, \bibinfo
		{author} {\bibfnamefont {M.}~\bibnamefont {Segev}}, \bibinfo {author}
		{\bibfnamefont {A.}~\bibnamefont {Szameit}},\ and\ \bibinfo {author}
		{\bibfnamefont {M.}~\bibnamefont {Heinrich}},\ }\bibfield  {title} {\bibinfo
		{title} {Fractal photonic topological insulators},\ }\href
	{https://doi.org/10.1126/science.abm2842} {\bibfield  {journal} {\bibinfo
			{journal} {Science}\ }\textbf {\bibinfo {volume} {376}},\ \bibinfo {pages}
		{1114} (\bibinfo {year} {2022})}\BibitemShut {NoStop}%
	\bibitem [{\citenamefont {Wu}\ and\ \citenamefont
		{Hu}(2015)}]{wu.prl.114.223901.2015}%
	\BibitemOpen
	\bibfield  {author} {\bibinfo {author} {\bibfnamefont {L.-H.}\ \bibnamefont
			{Wu}}\ and\ \bibinfo {author} {\bibfnamefont {X.}~\bibnamefont {Hu}},\
	}\bibfield  {title} {\bibinfo {title} {Scheme for achieving a topological
			photonic crystal by using dielectric material},\ }\href
	{https://doi.org/10.1103/PhysRevLett.114.223901} {\bibfield  {journal}
		{\bibinfo  {journal} {Phys. Rev. Lett.}\ }\textbf {\bibinfo {volume} {114}},\
		\bibinfo {pages} {223901} (\bibinfo {year} {2015})}\BibitemShut {NoStop}%
	\bibitem [{\citenamefont {Ferrando}\ \emph {et~al.}(2005)\citenamefont
		{Ferrando}, \citenamefont {Zacar\'es},\ and\ \citenamefont
		{Garc\'{\i}a-March}}]{ferrando.prl.95.043901.2005}%
	\BibitemOpen
	\bibfield  {author} {\bibinfo {author} {\bibfnamefont {A.}~\bibnamefont
			{Ferrando}}, \bibinfo {author} {\bibfnamefont {M.}~\bibnamefont
			{Zacar\'es}},\ and\ \bibinfo {author} {\bibfnamefont {M.-A.}\ \bibnamefont
			{Garc\'{\i}a-March}},\ }\bibfield  {title} {\bibinfo {title} {Vorticity
			cutoff in nonlinear photonic crystals},\ }\href
	{https://doi.org/10.1103/PhysRevLett.95.043901} {\bibfield  {journal}
		{\bibinfo  {journal} {Phys. Rev. Lett.}\ }\textbf {\bibinfo {volume} {95}},\
		\bibinfo {pages} {043901} (\bibinfo {year} {2005})}\BibitemShut {NoStop}%
	\bibitem [{\citenamefont {Klembt}\ \emph {et~al.}(2018)\citenamefont {Klembt},
		\citenamefont {Harder}, \citenamefont {Egorov}, \citenamefont {Winkler},
		\citenamefont {Ge}, \citenamefont {Bandres}, \citenamefont {Emmerling},
		\citenamefont {Worschech}, \citenamefont {Liew}, \citenamefont {Segev},
		\citenamefont {Schneider},\ and\ \citenamefont
		{H\"{o}fling}}]{klembt.nature.562.552.2018}%
	\BibitemOpen
	\bibfield  {author} {\bibinfo {author} {\bibfnamefont {S.}~\bibnamefont
			{Klembt}}, \bibinfo {author} {\bibfnamefont {T.~H.}\ \bibnamefont {Harder}},
		\bibinfo {author} {\bibfnamefont {O.~A.}\ \bibnamefont {Egorov}}, \bibinfo
		{author} {\bibfnamefont {K.}~\bibnamefont {Winkler}}, \bibinfo {author}
		{\bibfnamefont {R.}~\bibnamefont {Ge}}, \bibinfo {author} {\bibfnamefont
			{M.~A.}\ \bibnamefont {Bandres}}, \bibinfo {author} {\bibfnamefont
			{M.}~\bibnamefont {Emmerling}}, \bibinfo {author} {\bibfnamefont
			{L.}~\bibnamefont {Worschech}}, \bibinfo {author} {\bibfnamefont {T.~C.~H.}\
			\bibnamefont {Liew}}, \bibinfo {author} {\bibfnamefont {M.}~\bibnamefont
			{Segev}}, \bibinfo {author} {\bibfnamefont {C.}~\bibnamefont {Schneider}},\
		and\ \bibinfo {author} {\bibfnamefont {S.}~\bibnamefont {H\"{o}fling}},\
	}\bibfield  {title} {\bibinfo {title} {Exciton-polariton topological
			insulator},\ }\href {https://doi.org/10.1038/s41586-018-0601-5} {\bibfield
		{journal} {\bibinfo  {journal} {Nature}\ }\textbf {\bibinfo {volume} {562}},\
		\bibinfo {pages} {552} (\bibinfo {year} {2018})}\BibitemShut {NoStop}%
	\bibitem [{\citenamefont {Kartashov}\ and\ \citenamefont
		{Skryabin}(2017)}]{kartashov.prl.119.253904.2017}%
	\BibitemOpen
	\bibfield  {author} {\bibinfo {author} {\bibfnamefont {Y.~V.}\ \bibnamefont
			{Kartashov}}\ and\ \bibinfo {author} {\bibfnamefont {D.~V.}\ \bibnamefont
			{Skryabin}},\ }\bibfield  {title} {\bibinfo {title} {Bistable topological
			insulator with exciton-polaritons},\ }\href
	{https://doi.org/10.1103/PhysRevLett.119.253904} {\bibfield  {journal}
		{\bibinfo  {journal} {Phys. Rev. Lett.}\ }\textbf {\bibinfo {volume} {119}},\
		\bibinfo {pages} {253904} (\bibinfo {year} {2017})}\BibitemShut {NoStop}%
	\bibitem [{\citenamefont {Zhang}\ \emph
		{et~al.}(2023{\natexlab{b}})\citenamefont {Zhang}, \citenamefont
		{Bongiovanni}, \citenamefont {Wang}, \citenamefont {Wang}, \citenamefont
		{Xia}, \citenamefont {Hu}, \citenamefont {Song}, \citenamefont {Juki\'c},
		\citenamefont {Xu}, \citenamefont {Morandotti}, \citenamefont {Buljan},\ and\
		\citenamefont {Chen}}]{zhang.elight.3.5.2023}%
	\BibitemOpen
	\bibfield  {author} {\bibinfo {author} {\bibfnamefont {Y.}~\bibnamefont
			{Zhang}}, \bibinfo {author} {\bibfnamefont {D.}~\bibnamefont {Bongiovanni}},
		\bibinfo {author} {\bibfnamefont {Z.}~\bibnamefont {Wang}}, \bibinfo {author}
		{\bibfnamefont {X.}~\bibnamefont {Wang}}, \bibinfo {author} {\bibfnamefont
			{S.}~\bibnamefont {Xia}}, \bibinfo {author} {\bibfnamefont {Z.}~\bibnamefont
			{Hu}}, \bibinfo {author} {\bibfnamefont {D.}~\bibnamefont {Song}}, \bibinfo
		{author} {\bibfnamefont {D.}~\bibnamefont {Juki\'c}}, \bibinfo {author}
		{\bibfnamefont {J.}~\bibnamefont {Xu}}, \bibinfo {author} {\bibfnamefont
			{R.}~\bibnamefont {Morandotti}}, \bibinfo {author} {\bibfnamefont
			{H.}~\bibnamefont {Buljan}},\ and\ \bibinfo {author} {\bibfnamefont
			{Z.}~\bibnamefont {Chen}},\ }\bibfield  {title} {\bibinfo {title}
		{Realization of photonic $p$-orbital higher-order topological insulators},\
	}\href {https://doi.org/10.1186/s43593-022-00039-7} {\bibfield  {journal}
		{\bibinfo  {journal} {eLight}\ }\textbf {\bibinfo {volume} {3}},\ \bibinfo
		{pages} {5} (\bibinfo {year} {2023}{\natexlab{b}})}\BibitemShut {NoStop}%
	\bibitem [{\citenamefont {Bongiovanni}\ \emph {et~al.}(2024)\citenamefont
		{Bongiovanni}, \citenamefont {Hu}, \citenamefont {Wang}, \citenamefont
		{Wang}, \citenamefont {Jukić}, \citenamefont {Hu}, \citenamefont {Song},
		\citenamefont {Morandotti}, \citenamefont {Chen},\ and\ \citenamefont
		{Buljan}}]{bongiovanni.lpr.18.2400638.2024}%
	\BibitemOpen
	\bibfield  {author} {\bibinfo {author} {\bibfnamefont {D.}~\bibnamefont
			{Bongiovanni}}, \bibinfo {author} {\bibfnamefont {Z.}~\bibnamefont {Hu}},
		\bibinfo {author} {\bibfnamefont {Z.}~\bibnamefont {Wang}}, \bibinfo {author}
		{\bibfnamefont {X.}~\bibnamefont {Wang}}, \bibinfo {author} {\bibfnamefont
			{D.}~\bibnamefont {Jukić}}, \bibinfo {author} {\bibfnamefont
			{Y.}~\bibnamefont {Hu}}, \bibinfo {author} {\bibfnamefont {D.}~\bibnamefont
			{Song}}, \bibinfo {author} {\bibfnamefont {R.}~\bibnamefont {Morandotti}},
		\bibinfo {author} {\bibfnamefont {Z.}~\bibnamefont {Chen}},\ and\ \bibinfo
		{author} {\bibfnamefont {H.}~\bibnamefont {Buljan}},\ }\bibfield  {title}
	{\bibinfo {title} {$p$-orbital higher-order topological corner states in {2D}
			photonic {Su–Schrieffer–Heeger} lattices},\ }\href
	{https://doi.org/https://doi.org/10.1002/lpor.202400638} {\bibfield
		{journal} {\bibinfo  {journal} {Laser Photon. Rev.}\ }\textbf {\bibinfo
			{volume} {18}},\ \bibinfo {pages} {2400638} (\bibinfo {year}
		{2024})}\BibitemShut {NoStop}%
\end{thebibliography}
%apsrev4-2.bst 2019-01-14 (MD) hand-edited version of apsrev4-1.bst
%Control: key (0)
%Control: author (8) initials jnrlst
%Control: editor formatted (1) identically to author
%Control: production of article title (0) allowed
%Control: page (0) single
%Control: year (1) truncated
%Control: production of eprint (0) enabled
%

\end{document}